\documentclass[aps,prb,reprint,superscriptaddress]{revtex4-2}
\usepackage{graphicx}
\usepackage{amsfonts} 
\usepackage{dcolumn}% Align table columns on decimal point
\usepackage{bm}% bold math
\usepackage{braket}
\usepackage{comment}
\usepackage{alertmessage}
\usepackage{amsmath}
\usepackage[colorlinks,linkcolor=blue,citecolor=blue,urlcolor=blue]{hyperref}
\usepackage{orcidlink}
\usepackage{xr}
\begin{document}
	\title{Incommensurate-Stabilized Fractional Chern Insulator in Alternating Twisted Trilayer Graphene}

	\affiliation{
		State Key Laboratory of Semiconductor Physics and Chip Technologies, Institute of Semiconductors, Chinese Academy of Sciences, Beijing 100083, China
	}	
	\affiliation{
	Center for Quantum Matter, Zhejiang University, Hangzhou 310027, China
	}
	\affiliation{
		College of Materials Science and Opto-Electronic Technology, University of Chinese Academy of Sciences, Beijing 100049, China
	}

	\author{Moru Song\orcidlink{0009-0003-4842-6959}}
	
	\affiliation{
		State Key Laboratory of Semiconductor Physics and Chip Technologies, Institute of Semiconductors, Chinese Academy of Sciences, Beijing 100083, China
	}
	\affiliation{
	Center for Quantum Matter, Zhejiang University, Hangzhou 310027, China
	}
	\affiliation{
		College of Materials Science and Opto-Electronic Technology, University of Chinese Academy of Sciences, Beijing 100049, China
	}
	
	\author{Kai Chang\orcidlink{0000-0002-4609-8061}}
	\email{kchang@zju.edu.cn}
	\affiliation{
		Center for Quantum Matter, Zhejiang University, Hangzhou 310027, China
	}
	
	%\collaboration{}
	\date{\today}
	
	\begin{abstract}
		 Fractional Chern insulators (FCIs) typically emerge in topological flat bands and are regarded as lattice analogs of fractional quantum Hall states. Conventionally, the flat-band wavefunctions that support FCIs are expected to mimic the lowest Landau level, a condition that can be quantified by the quantum-geometric indicators. In realistic systems, however, FCIs often compete with lattice symmetry-breaking orders, especially when the hosting flat bands not ideal. \textcolor{black}{In this work, we propose stabilizing FCIs by exploiting the intrinsic incommensurability of alternating twisted trilayer graphene, which naturally suppresses competing charge-density-wave (CDW) phase while FCIs are less effected. Within an adiabatic approximation at the supermoir\'e scale, the effect of incommensuration on local physics can be quantified as phase shifts of interlayer coupling. Using exact diagonalization, we compute ground states in different local patches and uncover a strikingly counterintuitive result: the FCI gap increases as the quantum-geometric indicators worsen. Within certain parameter ranges, we further identify mixed phases where FCIs coexist with CDWs, but with CDWs confined only to patches of weak incommensurability. Finally, we provide experimental protocols and discuss how incommensuration enrich the system's topology and quantum geometry. Not only do our results establish incommensuration as a robust stabilizer of FCIs, but also provide a general paradigm for exploring strong-correlation physics in incommensurate systems.}
	\end{abstract}
	\maketitle
	
	\textit{Introduction.}---Fractional Chern insulators (FCIs) are topological orders enriched by translation symmetry, serving as lattice analogs of the fractional quantum Hall effect without the need for a magnetic field \cite{Regnault2011,Qi2011FQHEdual,Wu2012,Bernevig2012EmergentTLS,Liu2013,LiuZhao2022FCI}. They emerge in flat bands whose wavefunctions mimic lowest Landau level (LLL) behaviors, extending beyond the traditional paradigm of symmetry-broken phases. FCIs have been experimentally observed in moir\'e materials like twisted bilayer graphene (TBG), twisted transition metal dichalcogenides (TMDs) and multilayer graphene, where strong correlations and topology interplay to produce rich phase diagrams \cite{Spanton2018,YXie2021FCI,Cai2023,RMLFCI2024Julong}. In these systems, the quantum geometry of topological flat bands plays a central role in stabilizing FCIs. A key criterion is the fulfillment of the trace condition tied to vortexability, meaning that attaching a vortex to a Bloch state preserves it within the same band subspace \cite{Scaffidi2012FCIFQHE,Jackson2015,Ledwith_2023Vortexability}. In realistic moir\'e systems such ideal conditions are rarely met, which opens the door to competing symmetry-breaking orders \cite{LiuZhao2022FCI,PhysRevResearch.5.L012015}. \textcolor{black}{The most prominent competitor is the charge-density wave (CDW). A naive physical expectation then follows: CDW order is a symmetry breaking crystalline order, while the defining order of an FCI is topological and is less sensitive to the lattice background. When the two compete, incommensuration should suppress the crystalline channel more strongly than the topological one and thereby tilt the balance toward the FCI.}
	
	Motivated by this viewpoint, we introduce the concept of incommensuration stabilized fractional Chern insulators (I-FCIs).  In general, the incommensuration considered here is intrinsic and manifests as a cascade of length scales, from atomic ($\sim0.1\mathrm{nm}$) to moir\'e ($\sim10\mathrm{nm}$) and supermoir\'e ($\sim1\mu\mathrm{m}$) \cite{Zhu2020TTG,Mao2023SuperMoire,DevakulHTG2023,Guerci2024TopoFBTTG,Luo2024MoM,Xia2025Exp,SI}.
	\textcolor{black}{Within a continuum adiabatic framework, we quantify the effect of incommensuration on local physics by representing it as interlayer-coupling adiabatic phases. These phases vary slowly across the sample, allowing us to view the system as a mosaic of locally periodic moir\'e patches distinguished by their coupling phases}. The phases act as two emergent dimensions, closely analogous to a higher-dimensional Thouless pump \cite{Mao2023SuperMoire,Guerci2024TopoFBTTG}. Treating them as explicit parameters enables us to track, patch by patch, how intrinsic incommensuration reshapes band structures, quantum geometry, and interaction effects.

	\begin{figure*}
		\centering
		\includegraphics[width=0.8\linewidth]{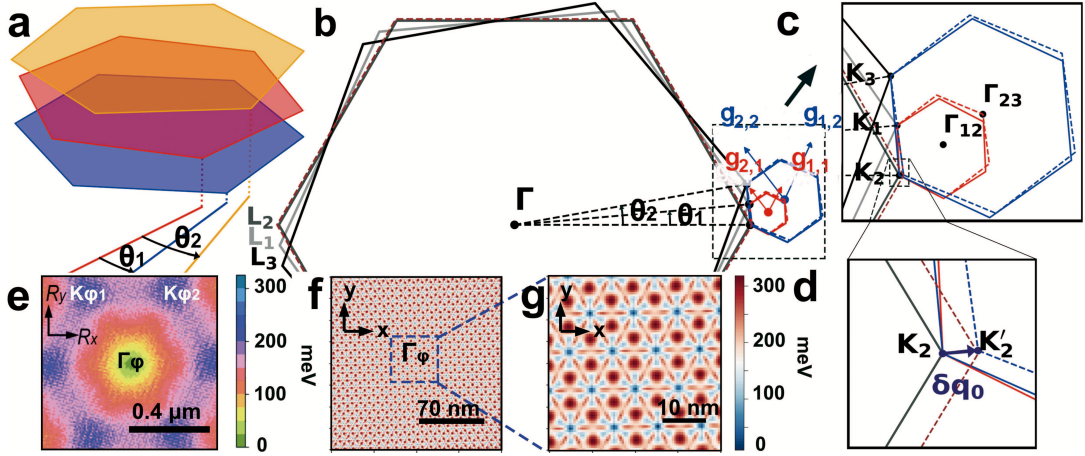}
		\caption{(a) Schematic of the A-TTG structure The (b,c) Moiré Brillouin zones (mBZs) for the two twist angles: each hexagon corresponds to the mini Brillouin zone of a single moiré lattice (red for $\theta_1$, blue for $\theta_2$). The lack of a common periodicity means there is no single unified mBZ. (d) Supermoir\'e scale scattering $\delta q_0$ with an expansion of the second layer graphene BZ (dashed dark red)  \textcolor{black}{(e) Validity of adiabatic Hamiltonian at $\mathbf{k}_\phi=\Gamma_\phi=(0,0)$ in supermoir\'e scale with $\Gamma_\phi$,$K_{\phi1}$ and $K_{\phi2}$ labels different patches.}  \textcolor{black}{(f, g) Distribution of the incommensurate moir\'e potential at different scale \cite{SI}}.  }
		\label{fig:ttgdigram}
	\end{figure*}
	
	Using alternating twisted trilayer graphene (A-TTG) as a platform \cite{Mora2019CTTG,Zhu2020TTG,Zhu2020ITTG,Nakatsuji2023RelaxTTG,AviramUri2023Mqc,Ezzi2024SuperMoire,Craig2024atomicTTG,Xia2025Exp}, we numerically identify that FCI are unexpectedly stabilized wheres local band geometries more deviated from LLL.
	%using exact diagonalization (ED) and particle entanglement spectrum (PES) analysis on different patches \cite{Regnault2015ES}, we show that, at the filling factor $\nu = 1/3$ for twist angles $\theta_1 = 1.4^\circ$ and $\theta_2 = 2.8^\circ$, the ground state locally exhibits FCI behavior. 
	\textcolor{black}{Specifically, we find that as incommensuration strengthens its influence on local physics, the FCI gap grows even while standard quantum-geometry indicators degrade, revealing a breakdown of the usual quantum-geometric prediction.} 
	This enhancement remains robust upon tuning $\kappa$, the parameter that controls interlayer coupling \cite{Tarnopolsky2018OrgMATBG}, and beyond a threshold near $\kappa \geq 0.6$ patches only weakly affected by incommensuration convert early into CDW order, producing a spatially mixed CDW-FCI regime with first-order character of phase transition on the supermoir\'e scale. Particle entanglement spectrum (PES) diagnostics identify the local FCIs as Laughlin-$1/3$ type. 
	\textcolor{black}{To account for the counterintuitive gap enhancement, a generalized trace-condition analysis for incommensurate systems is given however it still fail explaining the abnormal relationship. By tracking signatures of CDW, we conclude that intrinsic incommensuration chiefly suppresses the crystalline symmetry breaking channel more than FCI phases, allowing the topological order to prevail despite degraded trace condition.} 
	%\textcolor{black}{Incommensuration also have effect on global physics, the Wannier-center evolution records a loop along an emergent phase direction and the preferred charge localization across the supermoir\'e cell reverses, and a emergent $C_2$ number is numerically found when consider all patches into a dual 4D Hofstadter model.}
	Finally, we also discuss the experimental realization, higher dimensional topological enrichment of the second Chern number, and generalized geometric indicators in general incommensurate moir\'e materials.
	In summary, we identify intrinsic incommensuration within moir\'e stacks as a mechanism that stabilizes FCIs in flat bands depart from LLL geometry, and expand the landscape of FCI emergence across realistic incommensurate materials.

	\begin{figure*}
		\centering
		\includegraphics[width=\linewidth]{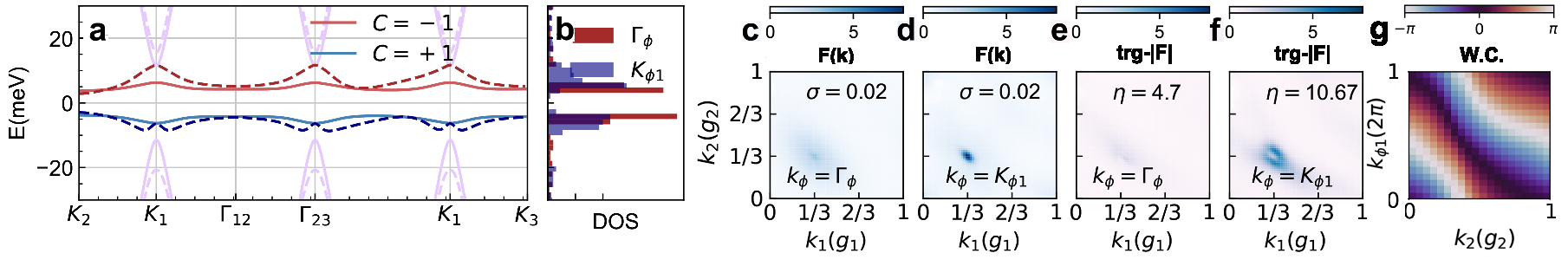}
		\caption{(a) Local band structure at $\kappa =0.6$, $\xi_v=-K$ for $\mathbf{k}_\phi=\Gamma_\phi$ and $\mathbf{k}_\phi= K_{\phi1}$ (solid/dashed) with hBN substrate $\Delta_\mathrm{hBN}=30~meV$, where supermoir\'e scale coordinate $\mathbf{k}_\phi$ with higher symmetries are $\Gamma_\phi=(0,0)$ and $K_{\phi 1,2}=\frac{2\pi}{3}(\pm1,\mp1)$ (See Fig. \ref{fig:ttgdigram}(b) and \ref{fig:edkappa} (a) insert). (b) Corresponding density of states (DOS). (c, d) Nearly uniform berry curvature for two valleys, the origin points is $K_2$. And we have $k_i=k_1g_{1i}+k_2g_{2i},~i=x,y$ (e, f) Trace condition for two locals where $\mathbf{g}_{i}=\mathbf{g}'_{i,1}$. (g) Wannier center (W.C.) one loop winding at $k_{\phi2}=0$ respect to both $k_2$ and $k_{\phi1}$ for valence flat bands.}
		\label{fig:ttgbandstructure}
	\end{figure*}
	\textit{Effect of incommensuration: an 1D moir\'e example---}\textcolor{black}{To build intuition, we begin with a concrete one-dimensional example with the Hamiltonian $\hat H=p^2/2m+V_1(x)+V_2(x)$. Take two periodic potential with periodicity $L_1=2\pi/G_1$, $L_2=2\pi/G_2$. Suppose $G_1=\pi$ and $G_2=3.14$; they share no common period, so there is no single unit cell and no single set of Bragg vectors that conserves momentum globally. 
	The remedy is to separate fast and slow scales. And we write the potential as
	$V_1(x)+V_2(x)=V_0\cos(G_1x)+V_0\cos(G_2x)=2V_0\cos\!\Big(\tfrac{G_1+G_2}{2}x\Big)\cos\!\Big(\tfrac{G_1-G_2}{2}x\Big),$
	which exposes a fast carrier and a slow envelope. The envelope varies on the long length $L_\phi=4\pi/|G_1-G_2|$, so we treat it adiabatically by introducing a slow coordinate $x\rightarrow R$ and a drifting phase $k_\phi=\delta q\cdot R$ with $\delta q=(G_1-G_2)/2$. Therefore we can substitute $V_0\rightarrow V_0^*(k_\phi)= V_0\cos\!\left(k_\phi\right)$. The single-particle problem becomes a family of locally periodic Hamiltonians $H(k_\phi)=p^2/2m+2V_0^*(k_\phi)\cos\!\Big(\tfrac{G_1+G_2}{2}x\Big)$. Thus, within the adiabatic approximation, the incommensurate effect is fully encoded as an intrinsic phase in the coupling constant of moir\'e potential. }
	
	\textit{Effect of incommensuration in A-TTG.—}The spirits are same in 2D cases. Here, we model the A-TTG structure in Fig.~\ref{fig:ttgdigram}(a) by starting from two coupled moir\'e Brillouin zones (mBZs) in Fig.~\ref{fig:ttgdigram}(b, c). The lengths of the two mBZs are $k_{\theta_I}= 8\pi/3a_0\sin (\theta_I/2)$ with $I=1,2$. Here $a_0=0.246~\mathrm{nm}$ is the graphene lattice constant, and the moir\'e reciprocal bases are $\mathbf{g}_{iI}=\mathcal{R}(\theta_I)\cdot \mathbf{b}_i-\mathbf{b}_i$ with $\mathcal{R}(\theta_I)$ the two-dimensional rotation matrix, where $\mathbf{b}_1= \frac{2\pi}{a_0}(1,-\sqrt{3}/3)$ and $\mathbf{b}_2= \frac{2\pi}{a_0}(1,\sqrt{3}/3)$ are the reciprocal bases of the middle graphene layer. Importantly, the incommensurate set of reciprocal lattice points $\mathcal{G}_\mathrm{ICM}:=\{\mathbf{G}\mid \sum_{i,I} n_{i,I} \mathbf{g}_{iI},~n_{i,I}\in\mathbb{Z}\}$ remains infinite even under a finite cutoff $|\mathbf{G}|<G_\mathrm{cut}$ (see Ref. \cite{Zhu2020TTG}). The effective continuum Hamiltonian of the $\pm K$ valleys reads
	\begin{gather}
		H_\mathrm{TTG}=
		\begin{bmatrix}
			v_F\,\boldsymbol{\sigma}_{\theta_1}^{\xi_v}\cdot \mathbf{k} & T_1(\mathbf{r}) & 0 \\
			T_1^\dagger(\mathbf{r}) & v_F\,\boldsymbol{\sigma}^{\xi_v} \cdot \mathbf{k} & T_2^\dagger(\mathbf{r}) \\
			0 & T_2(\mathbf{r}) & v_F\,\boldsymbol{\sigma}^{\xi_v}_{\theta_2}\cdot \mathbf{k} 
		\end{bmatrix},
		\label{eq:BM-mat}
	\end{gather}
	where $v_F=-8\times 10^5~\mathrm{m/s}$ is the Fermi velocity, $\xi_v=\pm1$ is the valley index, and $\boldsymbol{\sigma}_\theta^{\xi_v}=\mathcal{R}(\theta)\cdot (\xi_v\sigma_x,\sigma_y)$. The incommensuration enters through the interlayer matrices
	$
		T_I(\mathbf{r})=\sum_{s=0}^2 T_s^{\xi_v}\,e^{-i\,\mathbf{q}_{I,s}\cdot \mathbf{r}},
		\label{eq:InterlayerTmatICOM}
	$
	with $T_s^{\xi_v}=w\!\left(\kappa \sigma_0+\sigma_x\cos\!\frac{2\pi s}{3}+\xi_v\sigma_y\sin\!\frac{2\pi s}{3}\right)$, $w=110~\mathrm{meV}$, $\kappa=0.6$ the relaxation-induced corrugation, and $\mathbf{q}_{I,s}=\mathbf{K}_{L_I}-\mathbf{K}_{2}+\mathbf{g}_{I,s}$ the interlayer scattering vectors, where $\mathbf{g}_{I,0}:=\mathbf{0}$ and $\mathbf{K}_{L_I}$ with $L_{1,2}=1,3$ denote the $\pm K$ points of the outer layers. Following the sprint of  fast-slow separation, we perform a commensurate alignment by small reciprocal shifts $\delta\mathbf{q}_s$ so that the post-alignment vectors satisfy $\mathbf{q}'_{I,s}=\mathbf{q}_{I,s}+\delta\mathbf{q}_s$ and $|\mathbf{q}'_{1,0}|/|\mathbf{q}'_{2,0}|=p/q$ with $p,q\in\mathbb{Z}$~\cite{Mora2019CTTG,Mao2023SuperMoire}. To be specific, we slightly extend the second-layer BZ by $\delta\mathbf{q}_0$ as sketched in Fig.~\ref{fig:ttgdigram}(d), which yields $\delta\mathbf{q}_s=\mathcal{R}\!\left(\frac{2\pi s}{3}\right)\delta\mathbf{q}_0$. \textcolor{black}{We thereby quantify the effect of incommensuration on local physics as interlayer-coupling adiabatic phases by replacing $\delta\mathbf{q}_s\cdot \mathbf{r}\to \delta\mathbf{q}_s\cdot \mathbf{R}=k_{\phi s}$, where the slow variables $\mathbf{k}_\phi=(k_{\phi 1},k_{\phi 2})$ drift adiabatically on the supermoir\'e scale and supply two emergent dimensions~\cite{Mao2023SuperMoire}.} Consequently the interlayer blocks become
	\begin{gather}
		T_I(\mathbf{r},\mathbf{k}_\phi)= \sum_{s=0}^2 T_{s}^*(\mathbf{k}_\phi)\, e^{-i \mathbf{q}'_{I,s}\cdot \mathbf{r}},
		\label{eq:InterlayerTmatCOM}
	\end{gather}
	with $T_{s}^*(\mathbf{k}_\phi)= T_s^{\xi_v}\, e^{-i k_{\phi s}}$. At fixed $\mathbf{k}_\phi$ the problem is locally periodic, one can work with a finite plane-wave basis and a well-defined Bloch problem patch by patch, and then coherently stitch the solutions as $\mathbf{k}_\phi$ drifts with $\mathbf{R}$. \textcolor{black}{As shown in Fig.~\ref{fig:ttgdigram}(e), we compute the norm of difference between incommensurate and local commensurate moir\'e potential (at $\mathbf{k}_\phi=\Gamma_\phi=(0,0)$, see supplementary \cite{SI} S2 for definition), which characterize the validity of local Hamiltonian at $\Gamma_\phi$ in supermoir\'e scale. In general, the effective moir\'e potential for all patches and can be obtained by varying $\mathbf{k}_\phi$. In Supplementary Material \cite{SI}, we discuss the validation of such approximation in detail at different scale. This validity of patch work can be viewed directly from incommensurate moir\'e potential shown in Fig. \ref{fig:ttgdigram}(f, g), where the potential is locally periodic. This adiabatic construction turns intrinsic incommensuration into a quantitative organizing parameter that controls how local physics evolves across the supermoir\'e landscape.} 
	
	\begin{figure*}
		\centering
		\includegraphics[width=\linewidth]{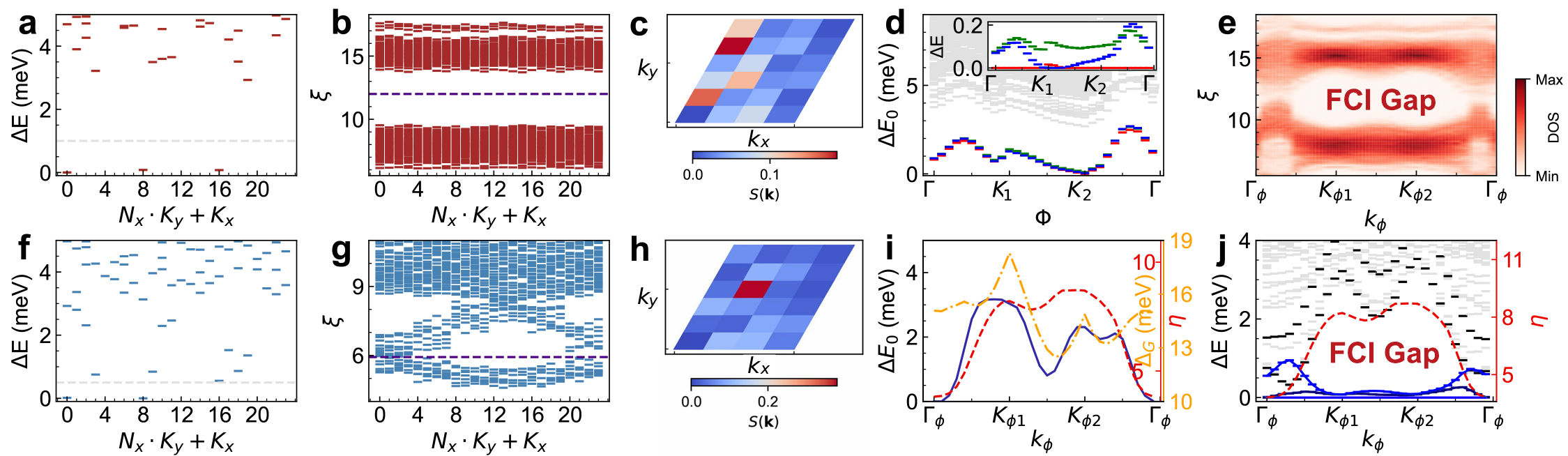}
		\caption{ED calculations for $N_x\times N_y = 4\times 6$ lattice with $8$ electron at $\kappa=0.6$. $\Delta E$ is relative energy to local ground states. $\Delta E_0$ is relative to minimum local ground states energy. $K_x$,$K_y$ are total momentum in the unit of $\mathbf{g}_{1,2}$. (a) FCI phase with 3-fold degenerated ground state at $\mathbf{k}_\phi=K_{\Phi1}$. (b) Corresponding PES where $1088$ total states under the gap for: $N_A= 3,N_B=4$. $e^{-\xi}$ labels eigenvalues of reduced density matrix. (c) Structural factor without CDW signature, where $k_x,k_y$ are momentum in reciprocal space. (d) Spectral flow of FCI phase along the loop pass high symmetry points in mBZ. (e) PES evolution with $\mathbf{k}_\phi$, showing FCI and CDW gap transition. (f) CDW ground state at $\mathbf{k}_\phi=\Gamma_\phi$. (g) CDW counting rule with $3C_8^3=168$ states under the gap. (h) Obvious CDW signature for Structural factor. \textcolor{black}{ (i) $\eta$ (red) and band gap towards remote band $\Delta_G$ (orange) variation with respect to Local ground states energy (purple). (j) Variation of FCI ground states (blue) and $\eta$ (red). }}
		\label{fig:edlocal}
	\end{figure*}	
	
	\textit{Local band structures and quantum geometries---}After solve the local band structure patch by patch, we then analyze the single-particle bands and introduce the quantum geometric indicators used below. As depicted in Fig.~\ref{fig:ttgbandstructure}(a), flat bands at two different patches are obtained (solid and dashed curve), which is obviously different. Here, an outer-layer hBN sublattice potential $H_\mathrm{hBN}=\Delta_\mathrm{hBN}\sigma_z/2$ is added with $(p,q)=(2,1)$. These flat bands carry first Chern numbers $C=\pm1$, and the gaps to remote bands are $\Delta_G\simeq 8\text{--}18~\mathrm{meV}$ [Fig.~\ref{fig:ttgbandstructure}(a,b)]. To quantify band geometry at fixed $\mathbf{k}_\phi$ we use the quantum geometric tensor $T_{\mu\nu}=\langle \partial_{k_\nu}u(\mathbf{k,k}_\phi) |\hat Q(\mathbf{k,k}_\phi)| \partial_{k_\mu}u(\mathbf{k,k}_\phi)\rangle$ with $Q(\mathbf{k,k}_\phi)=1-\lvert u(\mathbf{k,k}_\phi)\rangle\langle u(\mathbf{k,k}_\phi)\rvert$. Its real part $g_{\mu\nu}=\Re\,T_{\mu\nu}$ is the quantum metric and its imaginary part gives the Berry curvature $\mathcal{F}(\mathbf{k},\mathbf{k}_\phi)=2\,\epsilon^{\mu\nu}\Im[T_{\mu\nu}]$. We then define two band-integrated indicators at each $\mathbf{k}_\phi$, the standard derivation of berry curvature  $\sigma[\mathcal{F}](\mathbf{k}_\phi)$ and the violation of trace condition,
	\begin{gather}
		\eta[T](\mathbf{k}_\phi)=\int_{\mathrm{BZ}} d^2k\;\big[\operatorname{tr}g(\mathbf{k},\mathbf{k}_\phi)-|\mathcal{F}(\mathbf{k},\mathbf{k}_\phi)|\big],
	\end{gather}
	where a small $\sigma$ and a small $\eta$ signal proximity to the LLL and thus favor FCI stability~\cite{Shavit2024QuantumGeometryFCI}. Here, the Berry curvature is nearly uniform as indicated by $\sigma[\mathcal{F}]$ [Fig.~\ref{fig:ttgbandstructure}(c,d)], while the trace condition violation remains appreciable and provides a concrete baseline for comparison with interacting phases [Fig.~\ref{fig:ttgbandstructure}(c-f)]. 
	Additionally, as illustrate in Fig.~\ref{fig:ttgbandstructure}(e, f) and Fig. ~\ref{fig:edlocal}(a) $\mathbf{k}_\phi$ shows a monotonic increase of the trace-condition violation $\eta[T](\mathbf{k}_\phi)$.

	\textcolor{black}{Incommensuration can not only effect local physics heavily, but also play important roles to global topological structure. In particular, the Wannier center $\mathrm{W.C.}(k_2,\mathbf{k}_\phi) = \int d k_1 \bra{u_\mathbf{k}(\mathbf{k}_\phi)} i \partial_{k_1} \ket{u_\mathbf{k}(\mathbf{k}_\phi)}$ exhibits a loop winding along $k_{\phi1}$ across the supermoir\'e scale [Fig.~\ref{fig:ttgbandstructure}(g); see Supplementary Material \cite{SI} S8 for behavior along $k_{\phi2}$], indicating a topological Thouless pump that explicitly show the absence of a global translational symmetry.  A nontrivial second Chern number can also being found when we consider $\mathbf{k}_\phi$ as dual 4D Hofstadter model momentum \cite{SI}. Although no evidence currently in this system show the global topological properties has effects on local patches, however, it may plays an important role in experiments that need further investigations.}
	
	\textit{Signature of many body phases---}To begin with, we consider the protective Coulomb interaction of electrons at $\nu = 1/3$ filling of the lowest flat bands at $p/q = 2$ for the $-K$ valley. The local interacting Hamiltonian is given by:
	\begin{gather}
		H_I= \sum_{n_{1,2,3,4}}\sum_{{\mathbf{k}_{1,2,3,4}}} \bar{V}_{1,2,3,4} \eta_{n_1,\mathbf{k}_1}^\dagger\eta_{n_2,\mathbf{k}_2}^\dagger\eta_{n_3,\mathbf{k}_3}\eta_{n_4,\mathbf{k}_4},
		\label{eq:InteractHami}
	\end{gather}
	where $\eta^\dagger_{n,\mathbf{k}}$ create an electron at $\mathbf{k}$ of the $n$th band, 
	$ \bar{V}_{1,2,3,4} = \frac{1}{2} \delta'_{\mathbf{k}_1 + \mathbf{k}_2, \mathbf{k}_3 + \mathbf{k}_4}\sum_{\mathbf{G}} \tilde{V}(\mathbf{k}_1 - \mathbf{k}_4 + \mathbf{G}) \langle u_{n_1}(\mathbf{k}_1) | u_{n_4}(\mathbf{k}_4 - \mathbf{G}) \rangle \langle u_{n_2}(\mathbf{k}_2) | u_{n_3}(\mathbf{k}_3 + \mathbf{G}) \rangle$
	The dual gate Coulomb interaction is $\tilde{V}(\mathbf{q}) = \frac{e^2}{4 \pi \varepsilon G(\mathbf{k}_\phi)} \frac{2 \pi \tanh(\xi q / 2)}{q}$ \cite{Bernevig2021TBGIII,LiuZhao2022FCI},
	where $\varepsilon= 4\varepsilon_0$ is the dielectric constant of the material (see Supplementary Material \cite{SI} S11 for discussions on different $\varepsilon$), $G(\mathbf{k}_\phi)=N_e(\mathbf{k}_\phi)\sqrt{3}a_M^2/2\nu$ is the area of local region, which is in general non-uniform, $\xi=2a_M\sim20~ nm$ is the distance between the top and bottom gates and $a_M \approx qa_0/\theta_1$ is the local moir\'e lattice constant. Further numerical validation \cite{SI} show that remote bands is negligible for the relevant physics. Therefore, we use single-band projection results throughout. 
	
\begin{figure} 
		\centering
		\includegraphics[width=\linewidth]{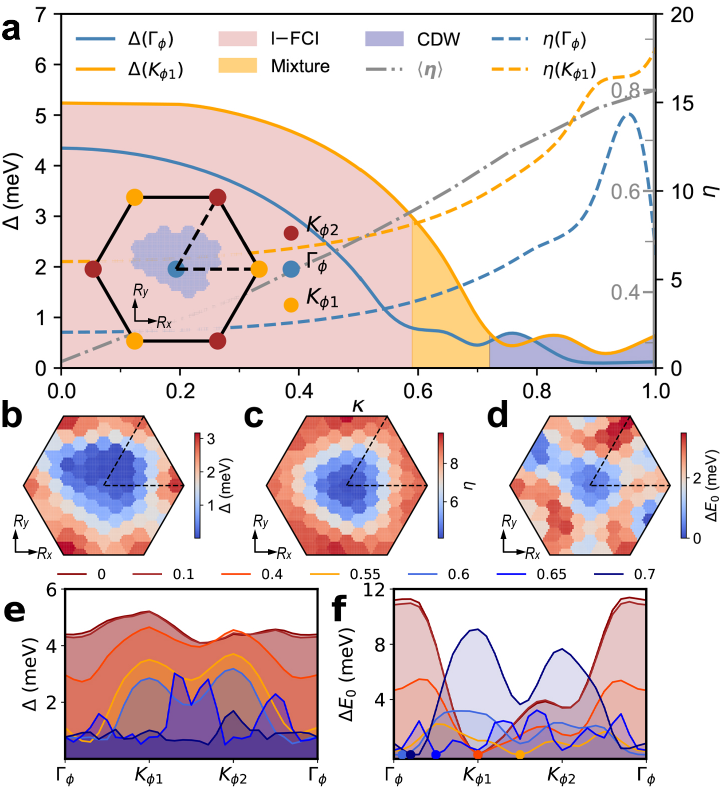}
		\caption{(a) Many body energy gap $\Delta$, single particle trace condition $\eta$ and mean value of generalized trace condition $\braket{\boldsymbol{\eta}}$ at different $\kappa$. Insert: Mixture phases distribution in Wagner Seitz supermoiré unit cell at $\kappa=0.6$, $\Gamma_\phi$ and $K_{\phi 1,2}$ labels higher local symmetries points at supermoiré scale in the real space.  The ED calculation is performed on the $10\times 10$ mesh in $\mathbf{k}_\phi$ space. (b) Many body gap distribution, \textcolor{black}{with $R_x,R_y$ are supermoir\'e scale coordinates.} (c) Distribution of $\eta$ (d) Ground state energy distribution. (e, f) Energy gap $\Delta$ and relative ground state energy $\Delta E_0$ distribution at different $\kappa$.}
		\label{fig:edkappa}
	\end{figure}

	As illustrate in Fig. \ref{fig:edlocal} and \ref{fig:edkappa}, We observed an FCI-CDW mixture phase at $\kappa=0.6$. For example, at $\mathbf{k}_\phi=K_{\phi 1}:=(2\pi,-2\pi)/3$ in Fig. \ref{fig:edlocal} (a), there are three fold degenerated ground states gaped about $3~meV$. Meanwhile, the spectral flow Fig. \ref{fig:edlocal} (d) shows that the system are always gaped under intersection of magnetic flux. \textcolor{black}{FCI states are identified through the PES, with the entanglement energies $\xi$, defined from the eigenvalues $e^{-\xi}$ of the reduced density matrix $\rho_A$ obtained by bipartitioning the system into $N_A=3$ and $N_B=5$ particles, serving as a fingerprint of topological order (see Supplementary Material \cite{SI} and Appendix B for details).} As shown in Fig. \ref{fig:edlocal} (b), states below the entanglement gap shows a clear signature of $(1,3)$-admissible counting rule \cite{JackPolynomials2008Haldane}. As the results, all these evidences provide strong signature of FCI states \cite{Regnault2011HaldaneStatics,Regnault2011,Bernevig2012MBTRS}. 
	
	\textcolor{black}{For comparison, at $\mathbf{k}_\phi = \Gamma_\phi := (0,0)$, the incommensurate effects locally vanish.} Here, Fig.~\ref{fig:edlocal}(f) exhibits a twofold degeneracy, and the PES shows a distinct signature. As illustrate in Fig. \ref{fig:edlocal} (g), the number of states in the lower branch are much fewer than those of the quasi-hole states, showing a typical CDW counting signature rather than FCIs \cite{Bernevig2012MBTRS}. Furthermore, we consider the structural factor 
	$
	S(\mathbf{k}) = \frac{1}{N_x N_y} \left( \langle \rho(\mathbf{k}) \rho(-\mathbf{k}) \rangle - (N_xN_y)^2 \delta_{\mathbf{k}, 0} \right)
	$. As illustrate in Fig. \ref{fig:edlocal} (h), $S(\mathbf{k})$ displays a pronounced peak, providing strong evidence for the presence of strip order and breaking of $C_3$ symmetry. As a compassion, $S(\mathbf{k})$ for FCI in Fig. \ref{fig:edlocal} (c) is featureless. \textcolor{black}{Additionally, we also compute  a regular commensurate case at another pairs of magic angles of $\theta_1=\theta_2$ at $\kappa=0.6$, which yields a similar CDW phase, verify the importance of incommensuration in Supplemental Material \cite{SI}.}

	The phase transition occurs at supermoir\'e scale as depicted in Fig. \ref{fig:edkappa} (a) insert. For example, in Fig. \ref{fig:edlocal} (e) phase transition occurs at $\mathbf{k}_\phi = \left(\pi, -\pi\right)/3$ and $\mathbf{k}_\phi = \left(-\pi, \pi\right)/3$, accompanied by the closing and reopening of the entanglement gap. As illustrate in Fig. \ref{fig:edkappa} (a), when $\kappa<0.6$ the system always hold local FCI phases with non-uniform many body gap $\Delta$. As $\kappa$ increase from $0.6$ to $0.7$, the CDW region grows and turns into full CDW phase when $\kappa>0.7$. Meanwhile, the exact trace condition $\eta=0$ is not well satisfied in whole parameter space as illustrate in Fig. \ref{fig:edkappa} (a). 
	
\begin{figure}
		\centering
		\includegraphics[width=\linewidth]{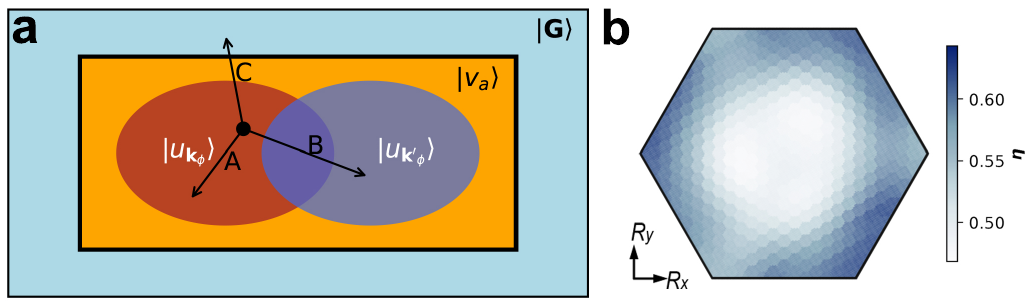}
		\caption{(a) Hierarchy of Hilbert space using different basis to describe. A: Attach vortex in local flat bands, representing satisfaction of traditional trace condition. B: Attach vortex that violet traditional one but satisfied by our new proposed trace condition. C: Violation of  both two trace conditions. (b) New trace condition numerical results in supermoire scale, representing leakage of vortices of C process.}
		\label{fig:vortexbilitydiagram}
	\end{figure}

	\textit{Incommensurate-stabilized FCI at supermoir\'e scale---}\textcolor{black}{A hallmark distinction of I-FCI is a counterintuitive geometry-stability correlation: as shown in Fig. \ref{fig:edlocal}(j), while the incommensurate influence on local physics strengthens, the trace-condition violation $\eta$ increases, yet the many-body FCI gap $\Delta$ grows along representative $\mathbf{k}_\phi$ lines and across full $\kappa$ sweeps, also see a comparison between Fig. \ref{fig:edkappa}(b, c). Although incommensurate itself may generalized the trace condition which we will introduced here after, such correlation still unchanged with the comparison of Fig. \ref{fig:edkappa}(c) and Fig. \ref{fig:vortexbilitydiagram}(b). Meanwhile, as shown in Fig. \ref{fig:edkappa}(e), this anomalous relationship is unchanged while $\kappa$ varies.   As shown in Fig. \ref{fig:edkappa}(a), another important fact is that, with increased $\kappa$, CDW order appears first near $\Gamma_\phi$ with smaller $\eta$. Thus, although larger $\eta$ would ordinarily bias a non-LLL-like flat band toward CDW in conventional settings, here it disfavors CDW and allows FCI gap $\Delta$ to increase. This paradox calls for a new explanation for why FCI is supported in such non-ideal local bands. Assuming the conventional quantum-geometry framework remains broadly valid, our results reveal an additional incommensuration-driven countervailing mechanism. In practice, non-ideal geometric indicators would ordinarily suppress FCIs, while FCIs compete with CDW order and CDW tends to from where incommensuration is weak. Since the only change to the Hamiltonian is the local phase induced by incommensuration, $ \mathbf{k}_\phi $, we infer that incommensuration suppresses crystalline (CDW) order, thereby offsetting geometric deterioration and stabilizing the local FCI.
	Consequently, I-FCI is enhanced in flat bands that deviate substantially from the LLL heuristic, unlike conventional FCIs that typically require near-ideal geometry.  }
	
	Another key feature of I-FCI lies in its emergent non-uniformity at the supermoir\'e scale. While our ED calculations fix the electron number at $N_e = 8$ locally, the actual ground state energy $E_0(\mathbf{k}_\phi)$ varies, indicating non-uniform density distribution [Fig.~\ref{fig:edkappa}(d)]. This can be understood via a self-consistent picture: regions with lower $E_0$ are energetically favorable. Thus, ED results for $E_0(\mathbf{k}_\phi)$ serve as a proxy for local electron preference. As shown in Fig.~\ref{fig:edkappa}(f), electrons favor $K_{\phi s}$ when $\kappa < 0.55$ and $\Gamma_\phi$ when $\kappa > 0.55$, consistent with the phase diagram in Fig.~\ref{fig:edkappa}(a), where the FCI phase prefers $K_{\phi s}$ regions and the CDW phase prefers $\Gamma_\phi$. As illustrated in Fig.~\ref{fig:edlocal}(i), this redistribution ultimately arises from the competition between band gap with remote band $\Delta_G$ and $\eta$.

	%\clearpage
	%\appendix
	%\setcounter{equation}{0}
	\textit{Experimental realization---}\textcolor{black}{Global transport experiment may meet difficulties at the first glace. In the coexistence regime, the two phases exhibit fundamentally different quantized Hall responses: the FCI carries a fractional Hall conductance, while the CDW exhibits an integer Hall conductance. The transport measurements should probe an average response over large sample regions, making the core signatures invisible.}
	
	\textcolor{black}{Nevertheless, we point out that the essential difference between FCI and CDW can still be resolved experimentally. A direct way is through the Streda formula (or Diophantine equation), $\frac{\partial \rho}{\partial B}=\sigma_H$, where $\rho$ is the electron density. Under a weak magnetic field, one can distinguish the two phases by their distinct slopes in Hall conductance: fractional versus integer. Concretely, the Landau fan diagram of longitudinal and Hall resistances would reveal two sets of lines---one with fractional slope and one with integer slope---that intersect at $B=0$. The coexistence of these intersecting fractional and integer features provides a clear transport signature of the incommensurate FCI-CDW mixed phase.}
	
	\textcolor{black}{Furthermore, for spectrum measurements, if a CDW-FCI coexistence occurs, local probes such us scanning tunneling microscopy (STM), single-electron transistor (SET) \cite{YXie2021FCI}, microwave impedance microscopy \cite{MIM2024} (MIM) or quantum twisting microscope (QTM) \cite{QuantumTwistingMicroscope2023} would most likely reveal CDW patterns and, in favorable cases, even CDW-FCI domain boundaries. However, this alone would still not conclusively establish the presence of an FCI.
	To make such identification more robust, one could combine local probes with a weak magnetic field. By scanning both electron density and magnetic field, one can analyze the evolution of local gaps via local density of states (LDOS) and determine whether the Landau fan slopes follow the Streda formula with fractional values. This would provide definitive evidence of a local FCI, and such an approach can naturally complement transport measurements.}
	
	\textit{Higher dimensional topological enrichment.---}The emergent phase $\mathbf{k}_{\phi}$ play a dual role: they set the real-space super-moiré length scale coordinate and, at the same time, enter the dual space as momentum. To motivate, the total Hamiltonian gives $H=\int G (\mathbf{k_\phi})d \mathbf{k_\phi} H(\mathbf{k_\phi})$, where $H(\mathbf{k_\phi})$ is given in Eq. \eqref{eq:BM-mat} and $G (\mathbf{k_\phi})$ is the area of local region. Here, we set $G(\mathbf{k_\phi})=\mathrm{const.}$ for simplicity.  Given $\alpha_{sj}=\mathbf{\Delta}_s\cdot \mathbf{g}_j$ and $\mathbf{\Delta}_s$ is the basis vector of middle layer graphene, we perform the Fourier transformation with $k_{ms}=\boldsymbol{\alpha}_s\cdot\mathbf{k}-k_{\phi s}\in[0,2\pi )$, yielding following dual Hamiltonian (see Supplementary Material \cite{SI} S8 for derivations),
	\begin{align}
		H_\mathrm{4D}
		&=\sum_{\gamma,\gamma'=1}^6\sum_{s=0}^2\sum_{\mathbf{n,m}}\left(\sum_{I=1}^2 t^{\gamma\gamma'}_{I,s} c^\dagger_{\mathbf{n}+\mathbf{e}_{Is},\mathbf{m},\gamma}c_{\mathbf{n},\mathbf{m},\gamma'}\right)
		\nonumber\\
		&+h_s^{\gamma\gamma'} e^{i\boldsymbol{\alpha}_s\cdot\mathbf{n}}c^\dagger_{\mathbf{n},\mathbf{m+1}_s,\gamma}c_{\mathbf{n},\mathbf{m},\gamma'}+h.c., 
		\label{eq:4DLatticeTheroy}
	\end{align}
	where $\mathbf{n,m}$ represent the dual space lattices. Here $\mathbf{n}$ is related to $\mathbf{G}\in \mathcal{G}$ in the original view of points. $\mathbf{1}_{0,1,2}=[0,0]_\mathbf{m},[1,0]_\mathbf{m},[0,1]_\mathbf{m}$ and $\mathbf{e}_{I 0,1,2}=[0,0]_\mathbf{n},[I,0]_\mathbf{n},[0,I]_\mathbf{n}$, with $t_{I,s}$ and $h_s$ are coupling matrix \cite{SI},  $c^\dagger_{\mathbf{n,m},\gamma} $ is the creation operator, with layer and sublattice degree of freedoms are labeled by $\gamma$. From Eq.~\eqref{eq:4DLatticeTheroy}, the first term can be interpreted as describing both nearest- and next-nearest-neighbor hoppings within the 
	$\mathbf{n}$-plane. In contrast, the second term corresponds to nearest-neighbor hoppings along the $\mathbf{m}$-plane, which acquire an additional phase that depends on the $\mathbf{n}$ coordinate. This structure effectively realizes a higher-dimensional analogue of the Hofstadter model \cite{AAH1980,Kraus2012AAHHarper,Kraus20134DQHE,Carr2020Dual}. The dual vector potential reads $\mathbf{A}=[0,0,\alpha_{11}x+\alpha_{12}y,\alpha_{21}x+\alpha_{22}y]$.
	
	This dual picture inspire us this system, in principle,  may facilitate the realization of the four-dimensional anomalous Hall effect under strong correlation \cite{SCZhang20014Dhall}. The second Chern number can be computed as $C_2 =+1$ \cite{Mochol2018SecondChern,YJXiu2023SecondChern} for the valence flat bands live in the manifold of the 4D torus parameterized by $k=(k_x,k_y,k_{\phi_1},k_{\phi_2})$ (see Supplementary Material \cite{SI}, S11). Note that  $G(\mathbf{k}_\phi)$ do not influence the global topological properties. Consequently, A-TTG can be considered topological equivalent to the lattice Hamiltonian Eq. (\ref{eq:4DLatticeTheroy}), validating the topological enrichment beyond traditional LLL picture.

	\textit{Generalized trace condition.---}\textcolor{black}{To assess whether the incommensuration-driven stabilization of FCIs is captured by quantum-geometric indicators, we analyze a general vortex attachment} \cite{Ledwith_2023Vortexability}. Here, we introduce a new type of trace condition based on the projector $\mathbf{Q}(k) = 1 - \sum_a |v_{\mathbf{k},a}\rangle\langle v_{\mathbf{k},a}|$, where $|v_{\mathbf{k},a}\rangle$ are orthogonal bases obtained from the QR decomposition of $|u_{\mathbf{k},\mathbf{k}_\phi}\rangle$ with respect to $\mathbf{k}_\phi$. The dimension along the $a$ direction is the rank of $|u_{\mathbf{k},\mathbf{k}_\phi}\rangle$ in the matrix representation under the plane wave bases, $d_a(\mathbf{k}) = \mathrm{Rank}\left(\mathcal{U}_{\mathbf{k}_\phi,\mathbf{G}}(\mathbf{k})\right) \sim 250$. Using these bases, the system can be fully quantized without need of introducing emergent dimension coordinate $\mathbf{k}_\phi$\cite{SI}.
	A quantum geometric tensor based on this new projector can be introduced as follows:
	\begin{gather}
		\mathbf{T}_{\mu\nu\mathbf{k}_\phi}(\mathbf{k}) = \langle \partial_{k_\nu} u_{\mathbf{k},\mathbf{k}_\phi} | \mathbf{Q}(\mathbf{k}) | \partial_{k_\mu} u_{\mathbf{k},\mathbf{k}_\phi} \rangle,
		\label{eq:QMG}
	\end{gather}
	Therefore, the new trace condition $\boldsymbol{\eta}_\mathbf{k_\phi} = \int_{\mathrm{mBZ}} d\mathbf{k} \left[ \mathrm{trRe}\left(\mathbf{T}\right) - \mathrm{Im}\left(\epsilon^{\mu\nu}\mathbf{T}_{\mu\nu}\right) \right]$ can be similarly defined using Eq.~(\ref{eq:QMG}). This process can be understood through Fig.~\ref{fig:vortexbilitydiagram}(a), where the local Bloch bases span an enlarged Hilbert subspace describing all local flat bands in the full system (see Appendix A for derivation of this argument). \textcolor{black}{Note that the value of $\eta$ naturally decrease for the shrink of Hilbert space represented by $\mathbf{Q}$.} The new trace condition implies that attaching vortices keeps the wavefunction within \textit{all moir\'e scale local flat bands}. Furthermore, $\boldsymbol{\eta}_\mathbf{k_\phi}$ increases with $\kappa$ as shown in Fig.~\ref{fig:edkappa}(a). \textcolor{black}{And we find the results in ~\ref{fig:vortexbilitydiagram}(b) shows a similar distribution with Fig.~\ref{fig:edkappa}(c), which means the generalized results still can not explain the abnormal relationship of trace condition and FCI gap. }

	\textcolor{black}{As complementary, this indicator can be applied to other incommensurate systems. Although currently, the distribution in Fig. \ref{fig:vortexbilitydiagram}(b) and Fig. \ref{fig:edkappa}(c) are similar, we do not rule out the possibility that in other systems incommensuration may directly affect FCI stability through modifications of quantum geometry.}
	
	\textit{Conclusion.---}In this letter, We introduce a general, adiabatic framework for analyzing strongly correlated phases in incommensurate moir\'e systems, by encoding incommensuration as a patch-dependent interlayer-coupling phase. \textcolor{black}{Within this framework, we uncover a robust and counter-intuitive effect: FCIs become more stable even as conventional quantum-geometry indicators deteriorate. This points to an intrinsic, incommensuration-driven stabilization mechanism that is distinct from the usual LLL picture: FCI is enhanced through the incommensuration induced suppression of CDW order.} Finally, we discuss the higher dimensional topological and geometric enrichment from incommensuration effect. Our work provide a new profound mechanism stabilized FCI phases and shed new light on exploring quasi-translational symmetry enriching  topological orders in general mutilayer-twistronics.

	\appendix
	\section{Vortexability for incommensurate system}
	The vortexability for Chern band is defined as \cite{Ledwith_2023Vortexability}:
	\begin{equation}
		\mathcal{Q}\hat{z}(\mathbf{r}) |\psi\rangle =0,
		\label{eq:Vortexbility}
	\end{equation}
	where \(\mathcal{Q} = 1-\sum_{ \mathbf{k}} |\psi_{\mathbf{k}}\rangle \langle \psi_{ \mathbf{k}}|\) is the projector out to the bands of interest. When \(\hat{z}(\mathbf{r}) = \hat{x} + i\hat{y}\), Eq.~\eqref{eq:Vortexbility} is equivalent to the trace condition \(\eta = 0\). Conversely, a nonzero \(\eta\) indicates the degree of violation of vortexability.  This can be demonstrate by giving another local projector \(Q(\mathbf{k}, \mathbf{k}_\phi) = 1 - |u_{\mathbf{k}, \mathbf{k}_\phi}\rangle \langle u_{\mathbf{k}, \mathbf{k}_\phi}|\), where \(|u_{\mathbf{k}, \mathbf{k}_\phi}\rangle = e^{-i\mathbf{k} \cdot \mathbf{r}} |\psi_{\mathbf{k}, \mathbf{k}_\phi}\rangle\) is the periodic Bloch function. 
	And \(|u_{\mathbf{k}, \mathbf{k}_\phi}\rangle\) can be expanded in the plane wave bases:$	|u_{\mathbf{k}, \mathbf{k}_\phi}\rangle = \sum_{\mathbf{G} \in \mathcal{G}} u_{\mathbf{G}, \mathbf{k}, \mathbf{k}_\phi} |\mathbf{G}\rangle$ \cite{SI}.
	
	Here, we define a new projector:
	$\boldsymbol{\mathcal{Q}} = 1 - \sum_{\mathbf{k}, a} |\chi_{\mathbf{k}, a}\rangle \langle \chi_{\mathbf{k}, a}|$, modifying vortexability to $\boldsymbol{\mathcal{Q}} \hat z ({\mathbf{r}}) |\psi_{\mathbf{k}, \mathbf{k}_\phi}\rangle = 0$. Therefore, $|\chi_{a,k}\rangle=\int d {k_\phi} \lambda_{a,k}(k_\phi)|\psi_k(k_\phi)\rangle$  are orthogonal bases: $\langle\chi_{a,p}|\chi_{b,k}\rangle=\delta_{a,b}\delta_{p,k}$. To be specific, we introduce the periodic orthogonal bases \(|v_{\mathbf{k}, a}\rangle = e^{-i\mathbf{k} \cdot \mathbf{r}} |\chi_{\mathbf{k}, a}\rangle\), we express them under plane wave expansions
	$|v_{\mathbf{k}, a}\rangle = \sum_{\mathbf{G} \in \mathcal{G}} v_{\mathbf{k}, \mathbf{G}, a} |\mathbf{G}\rangle$ where \(v_{\mathbf{k}, \mathbf{G}, a}\) are obtained via QR decomposition of matrix $[U(\mathbf{k})]_{\mathbf{G}, \mathbf{k}_\phi}=u_{\mathbf{G}, \mathbf{k}, \mathbf{k}_\phi}$ by desecrate sampling $\mathbf{k}_\phi$. The dimension of $a$ is large and depends on the cut-off of QR decomposition. The projector \(\mathbf{Q}(\mathbf{k}) = I - \sum_a |v_{\mathbf{k}, a}\rangle \langle v_{\mathbf{k}, a}|\) allows rewriting Eq.~\eqref{eq:Vortexbility} as:
	\begin{equation}
		\boldsymbol{\mathcal{Q}} \hat z ({\mathbf{r}}) |\psi_{\mathbf{k}, \mathbf{k}_\phi}\rangle = i e^{i\mathbf{k} \cdot \mathbf{r}} \left[ \mathbf{Q}(\mathbf{k}) \bar{\partial_{k}} |u_{\mathbf{k}, \mathbf{k}_\phi}\rangle \right],
	\end{equation}
	where $\bar{\partial_{k}}=\partial_{k_x}+i\partial_{k_y}$.
	Thus, vortexability requires \(\mathbf{Q}(\mathbf{k}) \bar{\partial_{k}} |u_{\mathbf{k}, \mathbf{k}_\phi}\rangle = 0\), linking it to the quantum geometric tensor and trace condition,
	\begin{align}
		&\boldsymbol{\eta}(\mathbf{k}_\phi)=\int d^2\mathbf{k} ~\mathrm{tr} \mathrm{Re}\mathbf{T}(\mathbf{k},\mathbf{k_\phi})-|\mathrm{Im}\mathbf{T}_{xy} (\mathbf{k},\mathbf{k_\phi})|\\
		&=\int d^2\mathbf{k} ~\bar\omega^\mu \mathbf{T}_{\mu\nu}(\mathbf{k},\mathbf{k_\phi})\omega^\nu 
		=\int d^2\mathbf{k} ~||\mathbf{Q}(\mathbf{k})\bar{\partial_k} u_\mathbf{k,k_\phi}||^2
		\nonumber,
		\label{eq:TraceCondition}
	\end{align}
	where $\boldsymbol{\omega}=[1,i]/\sqrt{2}$ \cite{SI}, and the modified trace condition for each local \(\mathbf{k}_\phi\) is then can be similar defined and denoted as \(\boldsymbol{\eta}_{\mathbf{k}_\phi}\) to describe the vortexability of our system numerically. 
	\section{Particle entanglement spectrum.}
	FCI states are identified through the PES, serving as a fingerprint of topological order \cite{Haldane2008ES,Regnault2011,Sterdyniak2011}. The ground states were expressed as \(\lvert \Omega_d \rangle = \sum_i e^{-\xi_i / 2} \lvert \Psi_i^A \rangle \otimes \lvert \Psi_i^B \rangle\), where \(e^{-\xi_i}\) and \(\lvert \Psi_i^A \rangle\) are the eigenvalues and eigenstates of the reduced density matrix \(\rho_A = \mathrm{Tr}_B(\rho)\), with the total density matrix defined by \(\rho = \frac{1}{3} \sum_d \lvert \Omega_d \rangle \langle \Omega_d \rvert\), see Supplementary Material \cite{SI} for more technique details. 
	PES was computed independently across momentum sectors \(\mathbf{K}_A\). The spectra were analyzed using $(k=1,r=3)$-admissible partition rules \cite{Regnault2011}, which restrict configurations to no more than one fermion per orbital and no more than $k$ fermions in $k+r$ consecutive orbitals. Configurations satisfying these criteria for \(N_A\) particles within an \(N_x \times N_y\) orbital mesh were enumerated. A PES exhibiting an entanglement gap, where the number of entanglement energies below the gap matches the number of valid configurations, provided strong evidence of the ground states residing in the FCI phase.
	
	\begin{acknowledgments}
		We acknowledge Zhengguang Lu for the helpful discussions on experimental realization. This work is supported by the National Natural Science Foundation of China (No. 12488101) and the Innovation Program for Quantum Science and Technology (Grant No. 2024ZD0300104).
	\end{acknowledgments}
	\bibliography{TTG_FCI_refs.bib}
\end{document}

\title{Supplementary Material for Incommensurate-Stabilized Fractional Chern Insulator in Alternating Twisted Trilayer Graphene}
	
	\affiliation{
		State Key Laboratory of Semiconductor Physics and Chip Technologies, Institute of Semiconductors, Chinese Academy of Sciences, Beijing 100083, China
	}
	\affiliation{
		Center for Quantum Matter, Zhejiang University, Hangzhou 310027, China
	}
	\affiliation{
		College of Materials Science and Opto-electronic Technology, University of Chinese Academy of Sciences, Beijing 100049, China
	}

	\author{Moru Song\orcidlink{0009-0003-4842-6959}}
	
	\affiliation{
		State Key Laboratory of Semiconductor Physics and Chip Technologies, Institute of Semiconductors, Chinese Academy of Sciences, Beijing 100083, China
	}
	\affiliation{
		Center for Quantum Matter, Zhejiang University, Hangzhou 310027, China
	}
	\affiliation{
		College of Materials Science and Opto-electronic Technology, University of Chinese Academy of Sciences, Beijing 100049, China
	}
	
	\author{Kai Chang\orcidlink{0000-0002-4609-8061}}
	\email{kchang@zju.edu.cn}
	\affiliation{
		Center for Quantum Matter, Zhejiang University, Hangzhou 310027, China
	}
	
	%\collaboration{}
	\date{\today}

	\maketitle
	
	\tableofcontents
	
	\section{Prove of arbitrary small shift vector}
	In the main text, we shift the two moire Brillouin zone to be commensurate. Here, we prove that this process can always be down without any mathematical defect at arbitrary twisted angles. In practice, when the system is commensurate, we have,
	\begin{gather}
		\mathbf{g}_{2i}=\sum_{j=1,2} \Lambda_{ij} \mathbf{g}_{1j}
		\label{eq:CommensurateCondition}
	\end{gather}
	with $\Lambda_{ij} \in \mathbb{Q}$ is a rational number and $\mathbf{g}_{Ii}$ is the moire reciprocal lattices for $I=1,2$ and $I=2,3$ layers. Generally, Eq. \ref{eq:CommensurateCondition} is always true if we slightly change it to $\Lambda_{ij}' \in \mathbb{R}$. The process we state in the main text can be mathematically turns into,
	\begin{gather}
		\delta \Lambda_{ij}=\Lambda_{ij}-\Lambda_{ij}'
	\end{gather} 
	And we have
	\begin{gather}
		\mathbf{g}_{2i}=\sum_j \Lambda_{ij}' \mathbf{g}_{1j}+ \delta \Lambda_{ij}\mathbf{g}_{1j}
	\end{gather}
	As the rational number is dense in the $\mathbb{R}$, for any $\Lambda_{ij}'$ we can find a series of  $\Lambda_{ij}$ that arbitrary close to it. As the result, the shift vector $\delta \mathbf{g}_i=\sum_j \delta \Lambda_{ij}\mathbf{g}_{1j}$ can be arbitrary small.
	\section{Validity of patch works}
	\begin{figure*}
		\centering
		\includegraphics[width=\linewidth]{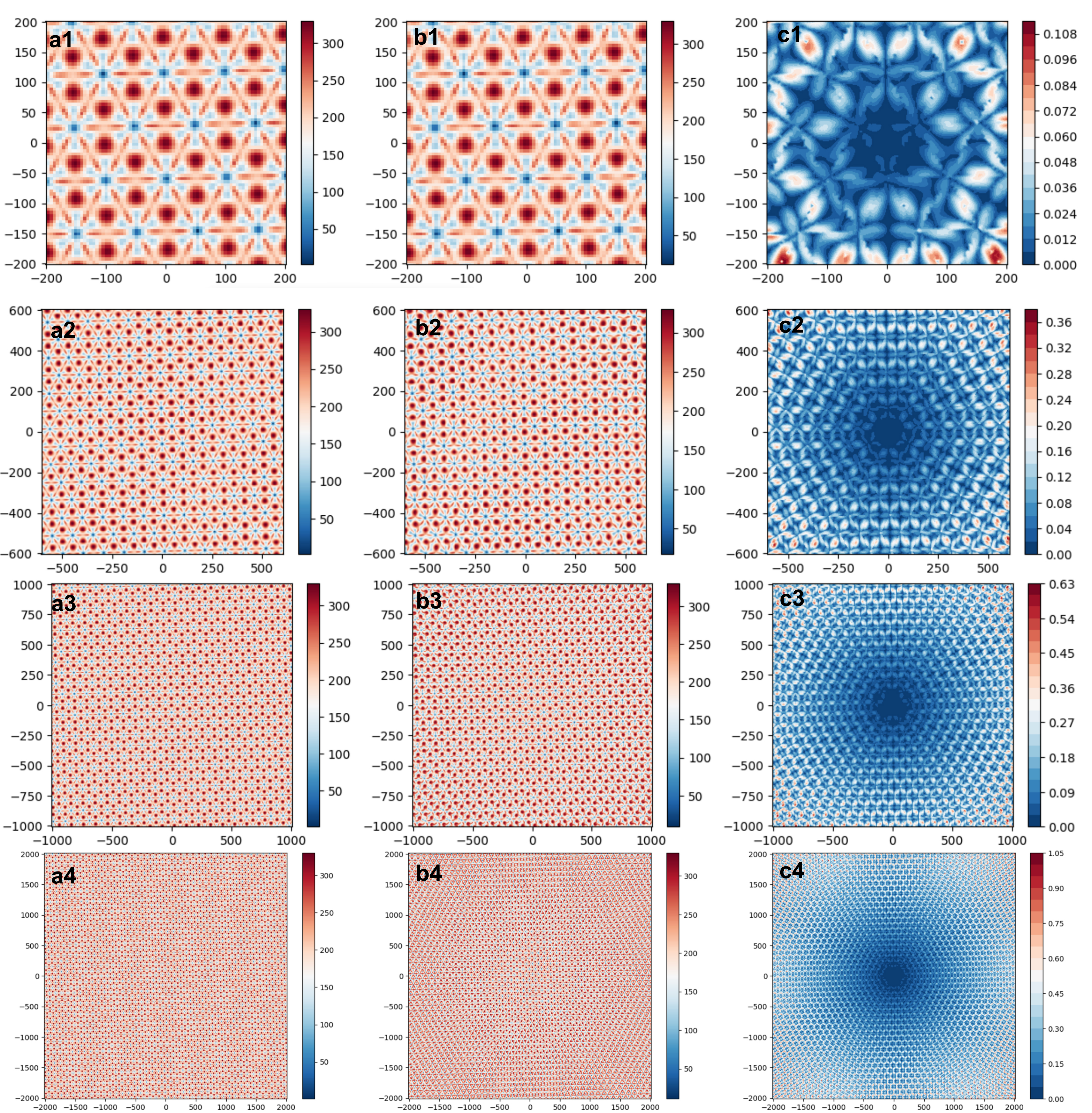}
	
		\caption{Validity of the adiabatic patch at $\mathbf{k}_\phi=0$. Unit of axis: $\AA$
			(a1–a4) Commensurate moiré potential $V_{\mathrm{com}}(\mathbf{r})$ around $\mathbf{r}=0$, shown over progressively larger square windows.
			(b1–b4) Corresponding incommensurate potential $V_{\mathrm{inc}}(\mathbf{r})$ evaluated at $\mathbf{k}_\phi=(0,0)$ over the same windows; for the smallest window [(a1,b1)] the two maps are nearly indistinguishable.
			(c1–c4) Relative-difference maps $\varepsilon(\mathbf{r}$ on the indicated windows, revealing a central “patch” where $\varepsilon\ll1$ and the adiabatic description by $H(\mathbf{k}_\phi=0)$ is accurate; $\varepsilon$ increases with distance from the origin.
		}
		\label{fig:incommvscommll}
	\end{figure*}
	\begin{figure*}
		\centering
		\includegraphics[width=\linewidth]{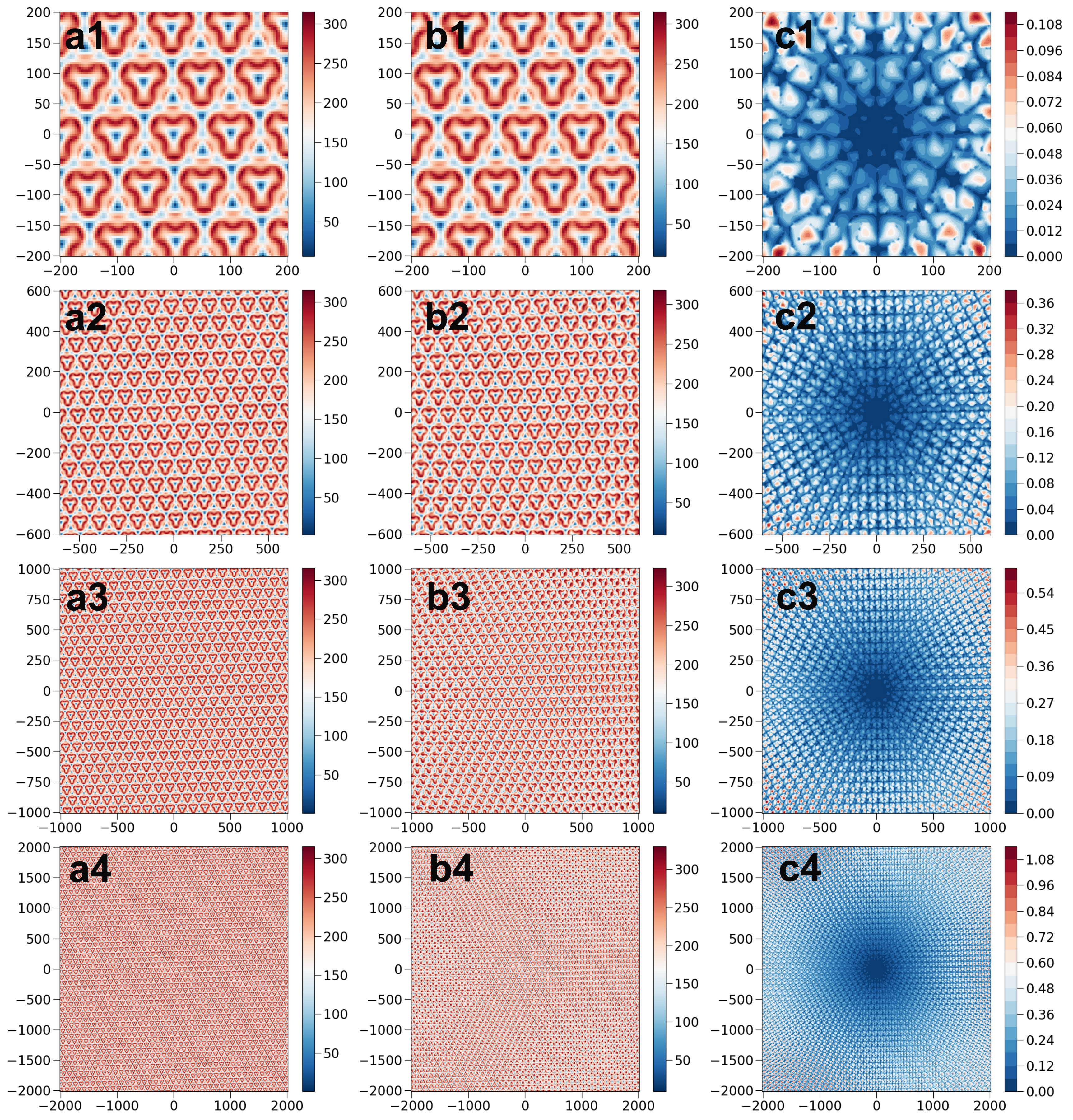}
		
		\caption{Validity of the adiabatic patch at $\mathbf{k}_\phi=K_{\phi1}$. Unit of axis: $\AA$
			(a1–a4) Commensurate moiré potential $V_{\mathrm{com}}(\mathbf{r})$ around $\mathbf{r}=0$, shown over progressively larger square windows.
			(b1–b4) Corresponding incommensurate potential $V_{\mathrm{inc}}(\mathbf{r})$ evaluated at $\mathbf{k}_\phi=(0,0)$ over the same windows; for the smallest window [(a1,b1)] the two maps are nearly indistinguishable.
			(c1–c4) Relative-difference maps $\varepsilon(\mathbf{r}$ on the indicated windows, revealing a central “patch” where $\varepsilon\ll1$ and the adiabatic description by $H(\mathbf{k}_\phi=0)$ is accurate; $\varepsilon$ increases with distance from the origin.
		}
		\label{fig:incommvscomm_Kphi1}
	\end{figure*}
	
	Here we visualize the commensurate and incommensurate moiré potentials and use them to assess the validity of the adiabatic local description $H(\mathbf{k}_\phi)$ at $\mathbf{k}_\phi=0$. In essence, the situation can be framed as follows: Hamiltonians $H(\mathbf{k}_\phi)$ defined on different local patches provide mutually inequivalent representations of the same system. No single local description can capture all global properties, but each can approximate the physics within its own neighborhood. On this basis, the task reduces to assessing, for each local Hamiltonian, how faithfully it represents the full system. We quantify this validity by the real-space distribution of the differences between the original moiré potential and its locally approximated counterpart. In regions where this differences is small, we regard the corresponding local Hamiltonian as an accurate and effective description of that region. 
	To make this point more transparent, one can compare with the eigenvalue distributions of the interlayer tunneling potential: $U(\mathbf{r})=|\text{eig}[\mathbf{T}(\mathbf{r})]_0|$ before and after the adiabatic approximation at $k_\phi=0$: 
	\begin{gather}
		\varepsilon(\mathbf{r})=\left|\frac{U_\text{ICM}(r)-U_\text{CM}(r,\mathbf{k}_\phi)}{w}\right|. 
	\end{gather}
	Note that in Fig. \ref{main-fig:ttgdigram}(b), we use ${U_\text{ICM}(r)-U_\text{CM}(r,\mathbf{k}_\phi=0)}$ for $w=110$ meV is just a constant. And here we only take the first eigenvalue of $\mathbf{T}(\mathbf{r})$.
	
	The incommensuration moire potential is given as:
	\begin{gather}
	\mathbf{T}(\mathbf{r})=
		\begin{bmatrix}
			0 & T_1(\mathbf{r}) & 0 \\
			T_1^\dagger(\mathbf{r}) &0 & T_2^\dagger(\mathbf{r}) \\
			0 & T_2(\mathbf{r}) &0
		\end{bmatrix},
		\label{eq:BM-mat}
	\end{gather}
	The incommensuration enters through the interlayer matrices
	\begin{gather}
		T_I(\mathbf{r})=\sum_{s=0}^2 T_s^{\xi_v}\,e^{-i\,\mathbf{q}_{I,s}\cdot \mathbf{r}},
	\end{gather}
	as defined in the main text.
	Their relative difference is extremely small in certain range of area compared with the moiré potential, confirming that the adiabatic treatment faithfully captures the effect of incommensuration (see Figure below). For example, when $ \varepsilon(\mathbf{r})$ under certain cut-off, i.e. $\sim 0.15$ we believe such adiabatic proximation is valid in such region.
	
	As shown in Fig. \ref{fig:incommvscommll}(a1–b1), over a small field of view the two potentials are visually indistinguishable, and the relative-difference map in Fig. \ref{fig:incommvscommll}(a3) is correspondingly tiny near the origin, growing gradually with distance. Upon enlarging the field of view [Fig. \ref{fig:incommvscommll}(a2–b2, a3–b3, a4–b4)], a central region emerges within which the commensurate approximation remains quantitatively accurate, consistent with the small maximal error summarized in Fig. \ref{fig:incommvscommll}(c1–c4). Beyond this region the incommensurate potential deviates increasingly from the periodic $\mathbf{k}_\phi=(0,0)$ potential, so the description by $H(\mathbf{k}_\phi=0)$ progressively degrades and eventually fails; in that regime one should switch to the Hamiltonian labeled by the appropriate local phase $\mathbf{k}_\phi$.
	
	Similarly, as shown in Fig. \ref{fig:incommvscomm_Kphi1}, we also depict the validity at $\mathbf{k}_\phi=K_{\phi1}$. The local structure is significantly different.

	\section{Geometric Criterion and vortexability}
	Geometry conditions have been identified as an important criterion for a model whether support an FCI phase, reflecting their adiabatic connection to fractional quantum hall states \cite{Qi2011FQHEdual,Scaffidi2012FCIFQHE}. This is usually related to how quantum geometry of the bands approximate that of the LLL, which quantified by the uniformity of berry curvature,
	$\sigma[\mathcal{F}](\mathbf{k}_\phi)=\sqrt{\int_{BZ} d^2\mathbf{k} \left(\mathcal{F}_{\mathbf{k}_\phi}(\mathbf{k})/2\pi-C\right)^2}$
	and the trace condition
	$\eta(\mathbf{k}_\phi)=\int_{BZ} d^2\mathbf{k} \left(\mathrm{tr}g_{\mathbf{k}_\phi}-|\mathcal{F}_{\mathbf{k}_\phi}|\right)$.
	
	To be specific, the trace condition is equivalent to the vortexability of single-particle Chern bands, which defined as \cite{Ledwith_2023Vortexability}:
	\begin{gather}
		\hat z(\mathbf{r}) |\psi\rangle= \mathcal{P} \hat z(\mathbf{r})|\psi\rangle
		\label{eq:Vortexbility}
	\end{gather}
	where $\mathcal{P}=\sum_\alpha |\psi_\alpha\rangle\langle\psi_\alpha|$ is the projector onto the bands of interest and $\alpha=n,\mathbf{k}$. If we only consider $\hat z(\mathbf{r})=\hat x+i\hat y$, then Eq. (\ref{eq:Vortexbility}) can be proved equivalent to the trace condition $\eta=0$ \footnote{ $\hat z(\mathbf{r})=\hat x+i\hat y +\delta(\mathbf{r})$ is beyond now giving trace condition, see Ref. \cite{Ledwith_2023Vortexability}  }. Alternatively speaking, the violation of trace condition suggesting the violation of vortexability. As the numerical results provided in the main text, if we only consider a specific local, the trace condition is violated greatly. However, in fact, if we consider the system globally, there are infinite number of flat bands. In this section we will review firstly on how trace condition turns to vortexability according to Ref. \cite{Ledwith_2023Vortexability} and back to our system in the end. To begin with, define $Q(\mathbf{k,k}_\phi)=1-|u_{\mathbf{k,k}_\phi}\rangle\langle u_{\mathbf{k,k}_\phi}|$, the geometric tensor is written,
	\begin{gather}
		T_{\mu\nu} (\mathbf{k},\mathbf{k_\phi}) = \langle\partial_{k_\nu} u_{\mathbf{k},\mathbf{k_\phi}}|Q(\mathbf{k,k}_\phi)|\partial_{k_\mu} u_{\mathbf{k},\mathbf{k_\phi}}\rangle,
	\end{gather}
	where $|u_{\mathbf{k},\mathbf{k_\phi}} \rangle =e^{-i \mathbf{k} \cdot \mathbf{r}}|\psi_{\mathbf{k},\mathbf{k_\phi}} \rangle$ is the periodic part of Bloch basis, in plane wave representation, it can be written as,
	\begin{gather}
		|u_{\mathbf{k},\mathbf{k_\phi}} \rangle= \sum_{\mathbf{G} \in \mathcal{G}} u_{\mathbf{G,k,k}_\phi } |\mathbf{G}  \rangle,
	\end{gather}
	with $u_{\mathbf{G,k,k}_\phi}$ is the eigenvector of target flat band from local Hamiltonian $H(\mathbf{k}_\phi)$. Note that, for we have choiosen an unified mBZ for the system, the plane wave basis themselves are independent of $\mathbf{k}_\phi$.
	The Fubini-Study metric $g_{\mu\nu}(\mathbf{k},\mathbf{k_\phi})=\mathrm{Re}T_{\mu\nu} (\mathbf{k},\mathbf{k_\phi})$ and the Berry curvature $\mathcal{F} (\mathbf{k},\mathbf{k_\phi})=\epsilon^{\mu\nu}\mathrm{Im}T_{\mu\nu}(\mathbf{k},\mathbf{k_\phi})$. The trace condition can therefore written as,
	\begin{align}
		\eta(\mathbf{k}_\phi)&=\int d^2\mathbf{k} ~\mathrm{tr} g(\mathbf{k},\mathbf{k_\phi})-\mathcal{F} (\mathbf{k},\mathbf{k_\phi})\\
		                     &=\int d^2\mathbf{k} ~\bar\omega^\mu T_{\mu\nu}(\mathbf{k},\mathbf{k_\phi})\omega^\nu \nonumber\\
		                     &=\int d^2\mathbf{k} ~||Q(\mathbf{k,k}_\phi)\bar{\partial_k} u_\mathbf{k,k_\phi}||^2
		                     \label{eq:TraceCondition}
	\end{align}
	where $\boldsymbol{\omega}=[1,i]^T/\sqrt{2}$ and $\bar{\partial_k}=\partial_{k_x}+i\partial_{k_y}$. Then, according to the Eq. (\ref{eq:Vortexbility}), we can write it equivalently with $\mathcal{Q}(\mathbf{k}_\phi)=1-\mathcal{P}=1-\sum_{\mathbf{k}}|\psi_{\mathbf{k,k}_\phi}\rangle\langle \psi_{\mathbf{k,k}_\phi}|$,
	\begin{gather}
		\mathcal{Q}(\mathbf{k}_\phi)\hat z(\mathbf{r})|\psi_\mathbf{k,k_\phi}\rangle = 0
		\label{eq:Vortexbility2}
	\end{gather}
	Therefore, as long as we have $\mathcal{Q}(\mathbf{k}_\phi) \hat z(\mathbf{r})|\psi_\mathbf{k,k_\phi}\rangle=0\Rightarrow Q({\mathbf{k},\mathbf{k_\phi}})\bar{\partial_k} |u_\mathbf{k,k_\phi}\rangle=0$ we can find the equivalence of trace condition and vortexability. To prove this relation, firstly, we take,
	\begin{gather}
		-i\nabla_\mathbf{k} |\psi_\mathbf{k,k_\phi}\rangle=\mathbf{\hat r} |\psi_\mathbf{k,k_\phi}\rangle+e^{i\mathbf{k}\cdot\mathbf{r}}(-i\nabla_\mathbf{k})|u_\mathbf{k,k_\phi}\rangle
		\label{eq:trialeq}
	\end{gather}
	Then we apply the projector $\mathcal{Q}(\mathbf{k}_\phi)$ to both sides of the equation. For $\mathcal{Q}(\mathbf{k}_\phi)\nabla_\mathbf{k} |\psi_\mathbf{k,k_\phi}\rangle=\mathcal{Q}(\mathbf{k}_\phi)(|\psi_\mathbf{k+dk,k_\phi}\rangle-|\psi_\mathbf{k,k_\phi}\rangle)/d\mathbf{k}$ is vanished, we have the equation,
	\begin{gather}
		\mathcal{Q}(\mathbf{k}_\phi)\mathbf{\hat r}|\psi_\mathbf{k,k_\phi}\rangle=	\mathcal{Q}(\mathbf{k}_\phi)ie^{i\mathbf{k}\cdot\mathbf{r}}\nabla_\mathbf{k}|u_\mathbf{k,k_\phi}\rangle
		\label{eq:Vortexbility3}
	\end{gather}
	Since \( \langle  \psi_{k',\mathbf{k}_\phi}| e^{i \mathbf{k} \cdot \mathbf{r}} \nabla_{\mathbf{k}} | u_{k,\mathbf{k}_\phi} \rangle \) vanishes for \( \mathbf{k} = \mathbf{k'} \) by translation symmetry, we have
	\begin{gather}
	\mathcal{Q}(\mathbf{k}_\phi) e^{i \mathbf{k} \cdot \mathbf{r}} \nabla_{\mathbf{k}} | u_{k,\mathbf{k}_\phi} \rangle = e^{i \mathbf{k} \cdot \mathbf{r}} Q(\mathbf{k},\mathbf{k}_\phi) \nabla_{\mathbf{k}} | u_{k,\mathbf{k}_\phi} \rangle,
	\end{gather}
	a rewriting of the right-hand side of Eq. (\ref{eq:Vortexbility3}). Multiplication by \(\boldsymbol{\omega} \) then yields
	\begin{gather}
	\mathcal{Q}(\mathbf{k}_\phi) z | \psi_{k,\mathbf{k}_\phi} \rangle = 0 \iff Q(\mathbf{k},\mathbf{k}_\phi) \bar{\partial_{k}} | u_{k,\mathbf{k}_\phi} \rangle = 0.
	\label{eq:TraceCondtion}
	\end{gather}
	From this derivation, we all use local operators at fixed $\mathbf{k}_\phi$, the Eq. (\ref{eq:TraceCondtion}) is not satisfied on both sides in our real A-TTG system. However,  Eq. (\ref{eq:TraceCondtion})  may too strong towards our system, for we allowing attaching vortex have not to stay in local bands. Therefore  we need to define a new projector 
	$\boldsymbol{\mathcal{Q}}=1-\sum_{\mathbf{k},a}|\chi_{\mathbf{k},a}\rangle\langle \chi_{\mathbf{k},a}|$. The modified vortexability then turns into
	$\boldsymbol{\mathcal{Q}} \mathbf{\hat r} | \psi_{k,\mathbf{k}_\phi} \rangle = 0$.
	Similarly, we introduce $|v_{\mathbf{k},a}\rangle=e^{-i\mathbf{k}\cdot\mathbf{r}}|\chi_{\mathbf{k},a}\rangle$ and 
	\begin{gather}
		|v_{\mathbf{k},a}\rangle=\sum_\mathbf{G\in\mathcal{G}} v_{\mathbf{k,G},a}|\mathbf{G}\rangle
	\end{gather}
	The basis $|v_{\mathbf{k},a}\rangle$ are complete orthogonal basis span the vector space spanned by $|u_{\mathbf{k,k}_\phi}\rangle$. This can be done by using QR decomposition. Note that the dimension of $a$ maybe $\mathbf{k}$ dependent. Similarly, we act $\boldsymbol{\mathcal{Q}}$ to both site of Eq. (\ref{eq:trialeq}), similar results we will obtain like Eq. (\ref{eq:Vortexbility3}), therefore we have,
	\begin{align}
		&\boldsymbol{\mathcal{Q}} \mathbf{\hat{r}} | \psi_{\mathbf{k, k_\phi}} \rangle = i \boldsymbol{\mathcal{Q}} \left[ e^{i \mathbf{k} \cdot \mathbf{\hat{r}}} \nabla_{\mathbf{k}} | u_{\mathbf{k, k_\phi}} \rangle \right] \\
		&= i \sum_{\mathbf{k}',a} \left( 1 - | \chi_{\mathbf{k}',a} \rangle \langle \chi_{\mathbf{k}',a} | \right) \left[ e^{i \mathbf{k} \cdot \mathbf{\hat{r}}} \nabla_{\mathbf{k}} | u_{\mathbf{k, k_\phi}} \rangle \right] \nonumber \\
		&= i  e^{i \mathbf{k} \cdot \mathbf{\hat{r}}} \nabla_{\mathbf{k}} | u_{\mathbf{k, k_\phi}} \rangle\nonumber \\
		&-\sum_{\mathbf{k}',a}  ie^{i \mathbf{k'} \cdot \mathbf{\hat{r}}} | v_{\mathbf{k'},a} \rangle \langle v_{\mathbf{k'},a} | e^{i (\mathbf{k-k}') \cdot \mathbf{\hat{r}}} \nabla_{\mathbf{k}} | u_{\mathbf{k, k_\phi}} \rangle \nonumber \\
		&=i  e^{i \mathbf{k} \cdot \mathbf{\hat{r}}} \left(\left(1-\sum_a | v_{\mathbf{k},a} \rangle \langle v_{\mathbf{k},a} |\right)\nabla_{\mathbf{k}} | u_{\mathbf{k, k_\phi}}\rangle \right)
		\label{eq:vortexablilty1}
	\end{align}
	Denote $\mathbf{Q}(k)=I-\sum_a|v_{\mathbf{k},a}\rangle\langle v_{\mathbf{k},a}|$. We can rewrite Eq. (\ref{eq:vortexablilty1}) as,
	\begin{gather}
		\boldsymbol{\mathcal{Q}} \mathbf{\hat{r}} | \psi_{\mathbf{k, k_\phi}} \rangle
		= i  e^{i \mathbf{k} \cdot \mathbf{\hat{r}}} \left[\mathbf{Q}(\mathbf{k})\nabla_\mathbf{k}|u_{\mathbf{k,k}_\phi}\rangle\right]
	\end{gather}
	As the results, the idea vortexability turns into the formula in the bracket to be zero. Consequently, using the fact $\mathbf{Q}(\mathbf{k})^2=\mathbf{Q}(\mathbf{k})$,  its norm square is directly corresponding to the quantum geometric tensor. 
	\begin{gather}
		\mathbf{T}_{\mu\nu}(\mathbf{k,k}_\phi) =  \langle\partial_{k_\nu} u_{\mathbf{k},\mathbf{k_\phi}}|\mathbf{Q}(\mathbf{k})|\partial_{k_\mu} u_{\mathbf{k},\mathbf{k_\phi}}\rangle,
		\label{eq:QGT}
	\end{gather}
	We can then use $\mathbf{T}_{\mu\nu}(\mathbf{k,k}_\phi)$ to compute the new trace condition for each local,
	\begin{gather}
		\boldsymbol{\eta}_\mathbf{k_\phi}=\int_{\mathrm{mBZ}} d\mathbf{k} ~ \mathrm{trRe}\left[\mathbf{T}\right]-\mathrm{Im}\left[\epsilon^{\mu\nu}\mathbf{T}_{\mu\nu}\right]
	\end{gather}
	This $\boldsymbol{\eta}_\mathbf{k_\phi}$ can be used to describe the vortexability of our system numerically.
    \section{Generalized Quantum Geometry }
    In last section, we have introduced the newly defined quantum metric, which will directly leading to the change of definition of connection and curvature. In principle, Eq. (\ref{eq:QGT}) can be witten use the the Non-abelian covariant derivatives $T_{\alpha \beta}^{\mu \nu} = \langle D^\mu u_\alpha | D^\nu u_\beta \rangle$ with $\alpha=\beta=k_\phi$ is exactly back to  Eq. (\ref{eq:QGT}) and,
    \begin{gather}
    	D^\nu| u_\beta \rangle= \partial^\nu |u_\beta\rangle - i\sum_a  A_{a \beta}^\nu|v_a\rangle,
    \end{gather}
    where $A_{a \beta}^\nu = i \langle v_a | \partial^\nu u_\beta \rangle$ is the complex non-abelian connection. Different from berry connection, $A_{a \beta}^{\nu*}\neq A_{a \beta}^\nu$. To be convenient, we use $\alpha,\beta$ to labels the emergent dimension coordinates, $a$ represent orthogonal basis, drop $\mathbf{k}$ and replace $\partial_{k_\mu} \rightarrow \partial_{\mu}$. The curvature field is defined through the anti-symmetric parts of $F_{\alpha\beta}=\epsilon_{\mu\nu} T_{\alpha \beta}^{\mu \nu}$. After some tedious derivation, one can obtain,
    \begin{align}
    	F_{\alpha\beta}^{\mu\nu} &= \sum_a \Big( \partial^\mu (\lambda_{a\alpha}^* A_{a\beta}^\nu) - \partial^\nu (\lambda_{a\alpha}^* A_{a\beta}^\mu) \Big) \label{eq:NonAbelianCurvature}
    	\\\nonumber
    	&- i \sum_a \Big( A_{a\alpha}^{\mu*} A_{a\beta}^\nu - A_{a\alpha}^{\nu *} A_{a\beta}^\mu \Big)    	
    \end{align}
    where $\lambda_{a\alpha}^*=\langle v_a|u_\alpha\rangle$. In general, Eq. (\ref{eq:NonAbelianCurvature}) is not gauge invariant. However, if we only consider $\alpha=\beta=\mathbf{k}_\phi$, Eq. (\ref{eq:NonAbelianCurvature}) is gauge invariant and can be written as,
    \begin{gather}
    	F_{\mathbf{k}_\phi}^{\mu\nu}=\mathcal{F}^{\mu\nu}_{\mathbf{k}_\phi}-i\sum_a \Big( A_{a \mathbf{k}_\phi}^{\mu*} A_{a \mathbf{k}_\phi}^\nu - A_{a \mathbf{k}_\phi}^{\nu *} A_{a \mathbf{k}_\phi}^\mu \Big)
    \end{gather}
    where the first term gives the berry curvature of system at local labeled by $\mathbf{k}_\phi$. The second term is in general non-vanishing for the non-hermitian properties of non-abelian connection $A_{a \mathbf{k}_\phi}^\nu$,the abelian berry connection can be represented as:
    \begin{gather}
    	\mathcal{A}^\nu_{\mathbf{k}_\phi}(\mathbf{k})=\sum_a \lambda^*_{a\mathbf{k}_\phi}(\mathbf{k}) A_{a\mathbf{k}_\phi}^\nu(\mathbf{k})
    \end{gather}
    In general, $F_{\mathbf{k}_\phi}^{\mu\nu}$ is not a Chern class for it is neither close nor exact. The integration of it is not quantized. However, one can also prove that $F_{\mathbf{k}_\phi}^{\mu\nu}$ is pure imagery as berry curvature. 
    
	\section{Quantization at supermoiré scale}
	In the main text, we treat $\mathbf{k}_\phi$ akin to an external classical field. The quantum operator can be viewed as the function of $\mathbf{k}_\phi$, $\eta^\dagger_{n,\mathbf{k}}(\mathbf{k}_\phi)$, and we assume $\{\eta_{n,\mathbf{k}},\eta^\dagger_{n',\mathbf{k'}}\}=\delta_{n,n'}\delta_\mathbf{k,k'}$ at a fixed local. The many body basis in Hilbert space is written in the single particle protective basis,
	$\ket{l,\mathbf{k}_\phi}=\prod_{\mathbf{k}\in l} \ket{u_\mathbf{k, k_\phi}}=\prod_{\mathbf{k}\in l} \eta^\dagger_{\mathbf{k}}(\mathbf{k}_\phi)\ket{0}$
	where $l$ is a specific assemble of orbitals of $N_e$ electrons. To obtain a full quantum theory of this system, we should note the non-orthogonal properties of local Bloch states and replace $\ket{u_\mathbf{k, k_\phi}}$ with $\ket{v_{\mathbf{k}, a}}$. Therefore we have
	\begin{align}
		\ket{l,\mathbf{k}_\phi}
		&=\prod_{\mathbf{k}_i\in l} \sum_{a_i} \lambda_{{a_i},\mathbf{k}_i}(\mathbf{k}_\phi)\ket{v_{\mathbf{k}_i, a_i}}\nonumber\\
		&=\sum_{\{a_i\}}\prod_{\mathbf{k}_i\in l}  \lambda_{{a_i},\mathbf{k}_i}(\mathbf{k}_\phi)\ket{v_{\mathbf{k}_i, a_i}}\nonumber\\
		&=\sum_{\{a_i\}}\prod_{\mathbf{k}_i\in l}  \lambda_{{a_i},\mathbf{k}_i}(\mathbf{k}_\phi)f^\dagger_{\mathbf{k}_i, a_i}\ket{0}\nonumber\\
		&=\sum_{\{a_i\}}\left(\prod_{\mathbf{k}_i\in l }\lambda_{{a_i},\mathbf{k}_i}(\mathbf{k}_\phi)\right)\prod_{\mathbf{k}_i\in l} f^\dagger_{\mathbf{k}_i, a_i}\ket{0}\nonumber\\
		&=\sum_{\mathcal{A}}\left(\prod_{\mathbf{k}_i\in l }\prod_{a_i\in \mathcal{A} }\lambda_{{a_i},\mathbf{k}_i}(\mathbf{k}_\phi)\right)\prod_{\mathbf{k}_i\in l} \prod_{a_i\in \mathcal{A} }f^\dagger_{\mathbf{k}_i, a_i}\ket{0}\nonumber\\
		&=\sum_{\mathcal{A}}C_{l \mathcal{A} }(\mathbf{k}_\phi)\ket{l,\mathcal{A}}
		\label{eq:manybodybasisTransfromation}
	\end{align}
	where $\sum_{\{a_i\}}:=\sum_{a_1,a_2,\cdots,a_{N_e}}$ with $i=1,\cdots,N_e$, $C_{l \mathcal{A} }=\prod_{\mathbf{k}\in l }\prod_{a\in \mathcal{A} }\lambda_{{a},\mathbf{k}}(\mathbf{k}_\phi)$ with $\mathcal{A}$ is a specific choice of $\mathcal{A}=(a_1,\cdots,a_{N_e})$, $\ket{l,\mathcal{A}}=\prod_{\mathbf{k}\in l}\prod_{a\in \mathcal{A}}f^\dagger_{\mathbf{k}, a}\ket{0}$ and $f^\dagger_{\mathbf{k}, a}$ is the creation operator satisfy,
	\begin{gather}
		\{f_{\mathbf{k}, a},f^\dagger_{\mathbf{k}', b}\}=\delta_{\mathbf{k,k'}}\delta_{ab}
	\end{gather} 
	This relation can naturally returns to the anti-commutator used in numerical ED calculation,
	\begin{align}
		\{\eta_{\mathbf{k}, \alpha},\eta^\dagger_{\mathbf{k}', \beta}\}
		&=\sum_{a,b}\lambda_{a,\alpha,\mathbf{k}}^*\lambda_{b,\beta,\mathbf{k'}}\{f_{\mathbf{k}, a},f^\dagger_{\mathbf{k}', b}\}\nonumber\\
		&=\delta_{\mathbf{k,k'}}\sum_{a}\lambda_{a,\alpha,\mathbf{k}}^*\lambda_{a,\beta,\mathbf{k'}}\nonumber\\
		&=\delta_{\mathbf{k,k'}} \sum_{a}\braket{u_{\mathbf{k},\alpha}|v_{\mathbf{k},a}}\braket{v_{\mathbf{k},a}|u_{\mathbf{k},\beta}}\nonumber\\
		&=\delta_{\mathbf{k,k'}}\braket{u_{\mathbf{k},\alpha}|u_{\mathbf{k},\beta}},
	\end{align}
	when $\alpha=\beta=\mathbf{k}_\phi$, we have $\braket{u_{\mathbf{k},\mathbf{k}_\phi}|u_{\mathbf{k},\mathbf{k}_\phi}}=1$
	
	Formally, the total protective Hamiltonian is also well defined, which is irreverent with local label $\mathbf{k}_\phi$,
	\begin{align}
		H&=\sum_{\mathbf{k},a,b}g_{\mathbf{k},a,b}f^\dagger_{\mathbf{k}, a}f_{\mathbf{k}, b}\nonumber\\
		&+\sum_{\{a\}}\sum_{\{\mathbf{k}\}} \tilde V_{\{\mathbf{k},a\}} f_{\mathbf{k}_1,a_1}^\dagger f_{\mathbf{k}_2,a_2}^\dagger f_{\mathbf{k}_3,a_3}f_{\mathbf{k}_4,a_4},
		\label{eq:FullHamiltonian}
	\end{align}
	where $g_{\mathbf{k},a,b}=\frac{1}{4\pi^2S_\mathrm{tot}}\int G(\mathbf{k}_\phi) d\mathbf{k}_\phi \lambda_{a}(\mathbf{k}_\phi)E_\mathbf{k}(\mathbf{k}_\phi)\lambda_{b}^*(\mathbf{k}_\phi)$ and $S_\mathrm{tot}$ is the total area of sample. Then, the interaction term become,
	\begin{align}
		\tilde V_{\{\mathbf{k},a\}}
		&=\frac{1}{4\pi^2S_\mathrm{tot}}\int G(\mathbf{k}_\phi) d\mathbf{k}_\phi\bar{V}_{\{\mathbf{k}\}}(\mathbf{k}_\phi)\lambda_{a_1}\lambda_{a_2}\lambda_{a_3}^*\lambda_{a_4}^*
		\nonumber\\
		&=\frac{1}{4\pi^2S_\mathrm{tot}}\int G(\mathbf{k}_\phi) d\mathbf{k}_\phi\frac{1}{2} \delta'_{\mathbf{k}_1 + \mathbf{k}_2, \mathbf{k}_3 + \mathbf{k}_4} 
		\nonumber\\
		&\times\braket{v_{\mathbf{k}_1a_1}|u_{\mathbf{k}_1}}\braket{v_{\mathbf{k}_2a_2}|u_{\mathbf{k}_2}}\braket{u_{\mathbf{k}_3}|v_{\mathbf{k}_3a_3}}\braket{u_{\mathbf{k}_4}|v_{\mathbf{k}_4a_4}}
		\nonumber\\
		&\times\sum_{\mathbf{G}} \tilde{V}(\mathbf{k}_1 - \mathbf{k}_4 + \mathbf{G}) \langle u_{\mathbf{k}_1} | u_{\mathbf{k}_4 - \mathbf{G}}\rangle  
		\langle u_{\mathbf{k}_2} | u_{\mathbf{k}_3 + \mathbf{G} + \delta \mathbf{G}} \rangle	\nonumber\\
		&=\frac{1}{4\pi^2S_\mathrm{tot}}\int G(\mathbf{k}_\phi) d\mathbf{k}_\phi\frac{1}{2} \delta'_{\mathbf{k}_1 + \mathbf{k}_2, \mathbf{k}_3 + \mathbf{k}_4} 
		\nonumber\\
		&\times\sum_{\mathbf{G}} \tilde{V}(\mathbf{k}_1 - \mathbf{k}_4 + \mathbf{G}) \braket{v_{\mathbf{k}_1a_1}|u_{\mathbf{k}_1}}\langle u_{\mathbf{k}_1} | u_{\mathbf{k}_4 - \mathbf{G}}\rangle
		\nonumber\\
		&\times\braket{u_{\mathbf{k}_4}|v_{\mathbf{k}_4a_4}}  
		\braket{v_{\mathbf{k}_2a_2}|u_{\mathbf{k}_2}}\langle u_{\mathbf{k}_2} | u_{\mathbf{k}_3 + \mathbf{G} + \delta \mathbf{G}} \rangle\braket{u_{\mathbf{k}_3}|v_{\mathbf{k}_3a_3}}
		\nonumber\\
		&=\frac{1}{4\pi^2S_\mathrm{tot}}\int G(\mathbf{k}_\phi) d\mathbf{k}_\phi\frac{1}{2} \delta'_{\mathbf{k}_1 + \mathbf{k}_2, \mathbf{k}_3 + \mathbf{k}_4} 
		\nonumber\\
		&\times\sum_{\mathbf{G}}\frac{e^2}{4 \pi \varepsilon G(\mathbf{k}_\phi)} \frac{2 \pi \tanh(\xi q / 2)}{q}\Big|_{q=|\mathbf{k}_1 - \mathbf{k}_4 + \mathbf{G}|}
		\nonumber\\
		&\times\bra{v_{\mathbf{k}_1a_1}}\hat U_{\mathbf{k}_1,\mathbf{k}_4-\mathbf{G}}(\mathbf{k}_\phi) \ket{v_{\mathbf{k}_4a_4}}
		\nonumber\\
		&\times\bra{v_{\mathbf{k}_2a_2}}\hat U_{\mathbf{k}_2,\mathbf{k}_3+\mathbf{G}+ \delta \mathbf{G}}(\mathbf{k}_\phi) \ket{v_{\mathbf{k}_3a_3}},
	\end{align}
	where $\hat U_{\mathbf{k}_1,\mathbf{k}_2}=\ket{u_{\mathbf{k}_1}}\braket{u_{\mathbf{k}_1}|u_{\mathbf{k}_2}}\bra{u_{\mathbf{k}_2}}$. Note that $G(\mathbf{k}_\phi)=N_e(\mathbf{k}_\phi)A_\mathrm{mBZ}/\nu$ represent the local Area and $N_e(\mathbf{k}_\phi)$ is the number of electrons for each local. The explicit value of $N_e(\mathbf{k}_\phi)$ should be self-consistent with the $\mathbf{k}_\phi$ independent condition of ground state energy $E_0(\mathbf{k}_\phi)=E_0$ (However, it is impossible to treat $N_e(\mathbf{k}_\phi)$ to self-consistent with fixed $E_0$ in real Exact diagonal calculation process). Then we have,
	\begin{align}
		\tilde V_{\{\mathbf{k},a\}}
		&=\frac{1}{4\pi^2S_\mathrm{tot}}\int d\mathbf{k}_\phi\frac{1}{2} \delta'_{\mathbf{k}_1 + \mathbf{k}_2, \mathbf{k}_3 + \mathbf{k}_4} 
		\nonumber\\
		&\times\sum_{\mathbf{G}}\frac{}{} \frac{2 \pi e^2 \tanh(\xi q / 2)}{4 \pi \varepsilon q}\Big|_{q=|\mathbf{k}_1 - \mathbf{k}_4 + \mathbf{G}|}
		\nonumber\\
		&\times\bra{v_{\mathbf{k}_1a_1}}\hat U_{\mathbf{k}_1,\mathbf{k}_4-\mathbf{G}}(\mathbf{k}_\phi) \ket{v_{\mathbf{k}_4a_4}}
		\nonumber\\
		&\times\bra{v_{\mathbf{k}_2a_2}}\hat U_{\mathbf{k}_2,\mathbf{k}_3+\mathbf{G}+ \delta \mathbf{G}}(\mathbf{k}_\phi) \ket{v_{\mathbf{k}_3a_3}},
	\end{align}
	\begin{figure*}
		\centering
		\includegraphics[width=\linewidth]{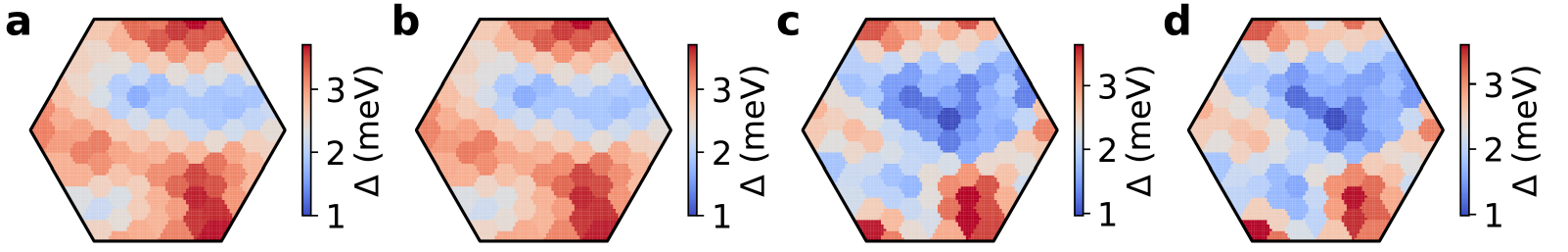}
		\caption{(a) Single band projection or involving consider conduction flat band at $4\times3$ lattice. (b) Consider remote flat band with at most one electron filled. (c) Consider remote full filled Dirac band. (d) Consider remote one hole maximum filled Dirac band.}
		\label{fig:manybodygapconvergence}
	\end{figure*}
	This enable us to consider the many body ground states on two sides.
	On the one hand, we can obtain degenerated local ground states (with degeneracy i.e.,$d(\mathbf{k}_\phi)=3$ for local FCI at $\nu=1/3$) numerically, and we denote them as
	\begin{align}
		\ket{\Omega_d(\mathbf{k}_\phi)}
		&= \sum_l \Omega_{dl}(\mathbf{k}_\phi)\ket{l,\mathbf{k}_\phi} \nonumber\\
		&= \sum_l \sum_{\mathcal{A}}\Omega_{dl}(\mathbf{k}_\phi)C_{l \mathcal{A} }(\mathbf{k}_\phi)\ket{l,\mathcal{A}},
		\label{eq:GSlocal}
	\end{align}
	where Eq. (\ref{eq:manybodybasisTransfromation}) is used to transform local ground states into the orthogonal basis representation. On the another hand, we can formally write down the ground states of the full system (Eq. (\ref{eq:FullHamiltonian})), which is independent of $\mathbf{k}_\phi$
	\begin{align}
		\ket{\Omega^0_{d_0}}=\sum_l\sum_{\mathcal{A}}\Omega^0_{{d_0},l,\mathcal{A}}\ket{l,\mathcal{A}}
		\label{eq:GSunique}
	\end{align}
	In principle, we can carry out another QR decomposition to Eq. (\ref{eq:GSlocal}) and find the degeneracy of $d_0$ in Eq. (\ref{eq:GSunique}). Denote $D=\{d,\mathbf{k}_\phi\}$ and $\mathcal{B}=(l,\mathcal{A})$, we can decompose the matrix $\boldsymbol{\Omega}_{D,\mathcal{B}}=\Omega_{dl}(\mathbf{k}_\phi)C_{l \mathcal{A} }(\mathbf{k}_\phi)$ and obtain,
	\begin{gather}
		\ket{\Omega^0_{d_0}}=\sum_{D} \Lambda_{d_0,D}\ket{\Omega_D}
	\end{gather}
	Although Eq. (\ref{eq:GSlocal}) can be obtained approximately in a small cluster of electrons locally, it is still impossible to obtain Eq. (\ref{eq:GSunique}) numerically, for handle such giant Hilbert space in practice is impossible: Typically, QR decomposition shows that the dimension along $a$ is about $d_a\sim250$ and the dimension of Hilbert space is,
	\begin{gather}
		\mathrm{dim}\mathcal{H}=\sum_{l_i}^{C_{N_xN_y}^{N_o}} \left(\prod_{\mathbf{k}_j\in l_i} 2^{d_{aj}(\mathbf{k}_j)}\right)
	\end{gather} 
	Note that $N_o$ here is the number of occupied orbitals in momentum degree of freedom rather than the total number of electron. $\mathbf{k}_\phi$ plays the role that smearing the $a$ degree of freedom by Eq. (\ref{eq:manybodybasisTransfromation}), so the total number of electrons is not specifically chosen.
	Consequently, we have proved that this system is well-defined in quantum at supermoiré scale Eq. (\ref{eq:FullHamiltonian}).
	
	\section{Particle entanglement spectrum}
	To determine the topological phase of the ground states in our fractional Chern insulator (FCI) system, we utilized the particle entanglement spectrum (PES) as a diagnostic tool, recognized for its ability to fingerprint topological order and indicate quasiparticle excitations \cite{Haldane2008ES,Regnault2011,Sterdyniak2011}. The ground states were expressed as \(\lvert \Omega_d \rangle = \sum_i e^{-\xi_i / 2} \lvert \Psi_i^A \rangle \otimes \lvert \Psi_i^B \rangle\), where \(e^{-\xi_i}\) and \(\lvert \Psi_i^A \rangle\) are the eigenvalues and eigenstates of the reduced density matrix \(\rho_A = \mathrm{Tr}_B(\rho)\), with the total density matrix defined by \(\rho = \frac{1}{3} \sum_d \lvert \Omega_d \rangle \langle \Omega_d \rvert\). 
	
	The states \(\lvert \Psi_i^A \rangle\) were constructed block-diagonally within the bases \(\lvert l_A \rangle = \prod_{\mathbf{k} \in l_A} \eta_{k}^\dagger \lvert 0 \rangle\), representing configurations of \(N_A\) particles occupying orbital labels \(\mathbf{k}\) with total momentum \(\mathbf{K}_A = \sum_{\mathbf{k} \in l_A} \mathbf{k}\). For each ground state sector with total momentum \(\mathbf{K}_d\), the complementary momentum \(\mathbf{K}_B = \mathbf{K}_d - \mathbf{K}_A\) defined the bases \(\lvert l_B \rangle\) for \(N_B = N_e - N_A\) particles. In fermionic systems, the combined bases states were given by \(\lvert l \rangle = (-1)^{s_{AB}} \lvert l_A \rangle \otimes \lvert l_B \rangle\), where \(s_{AB}\) accounts for the fermionic sign from the ordering of orbital labels \(\{k_1^A, \ldots, k_{N_A}^A, k_1^B, \ldots, k_{N_B}^B\}\), with each orbital label \(k = N_x k_y + k_x\) enforcing the normal ordering \(\eta_{k_1}^\dagger \cdots \eta_{k_{N_e}}^\dagger \eta_{k_{N_e}} \cdots \eta_{k_1}\) for \(k_1 < k_2 < \ldots < k_{N_e}\). The reduced density matrix elements were calculated as \([\rho_{A}]_{ij} = \sum_{l_B} (-1)^{(s_{A_iB}+s_{A_jB})} \Omega_{\{l_{A_i}\otimes l_B\}}^* \Omega_{\{l_{A_j}\otimes l_B\}}\), where \(\lvert \Omega \rangle = \sum_l \Omega_l \lvert l \rangle\).
	
	PES was computed independently across momentum sectors \(\mathbf{K}_A\), with \(K_{Ax}\) ranging from 0 to \(N_x\) and \(K_{Ay}\) from 0 to \(N_y\). The spectra were analyzed using $(k=1,r=2)$-admissible partition rules \cite{Regnault2011}, which restrict configurations to no more than one fermion per orbital and no more than $k$ fermions in $k+r$ consecutive orbitals. bases states \(\lvert l_v \rangle\) satisfying these criteria for \(N_A\) particles within an \(N_x \times N_y\) orbital mesh were enumerated. A PES exhibiting an entanglement gap, where the number of entanglement energies below the gap matches the number of valid bases states \(\lvert l_v \rangle\), provided strong evidence of the ground states residing in the FCI phase.
	
	\section{Validity of single band projection at $4\times 6$ Lattice}
	\begin{figure}
		\centering
		\includegraphics[width=\linewidth]{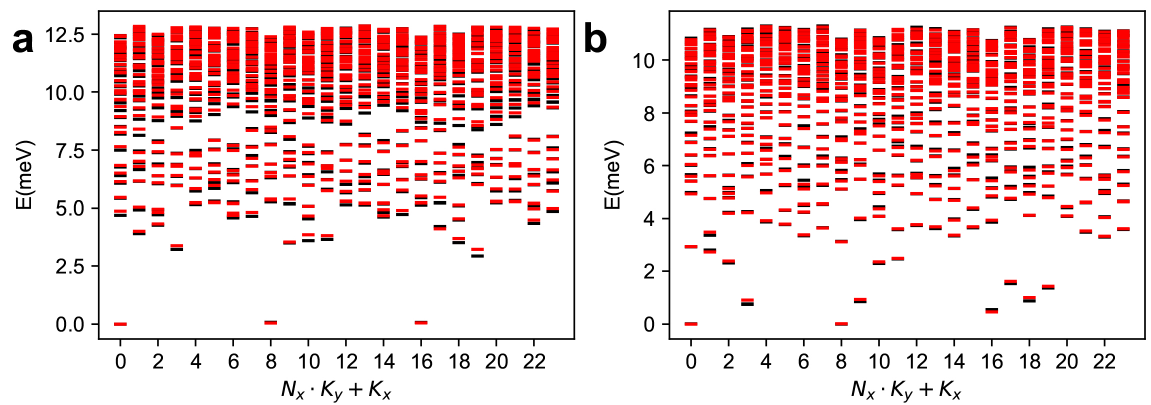}
		\caption{Effect when consider full filled Dirac band (red) and single band projection (black) at $N_x,N_y=4,6$ and $\kappa=0.6$, (a) $\mathbf{k}_\phi=K_{\phi 1}$, (b)$ \mathbf{k}_\phi=G_{\phi }$.}
		\label{fig:convegence}
	\end{figure}
	\begin{figure}
		\centering
		\includegraphics[width=1\linewidth]{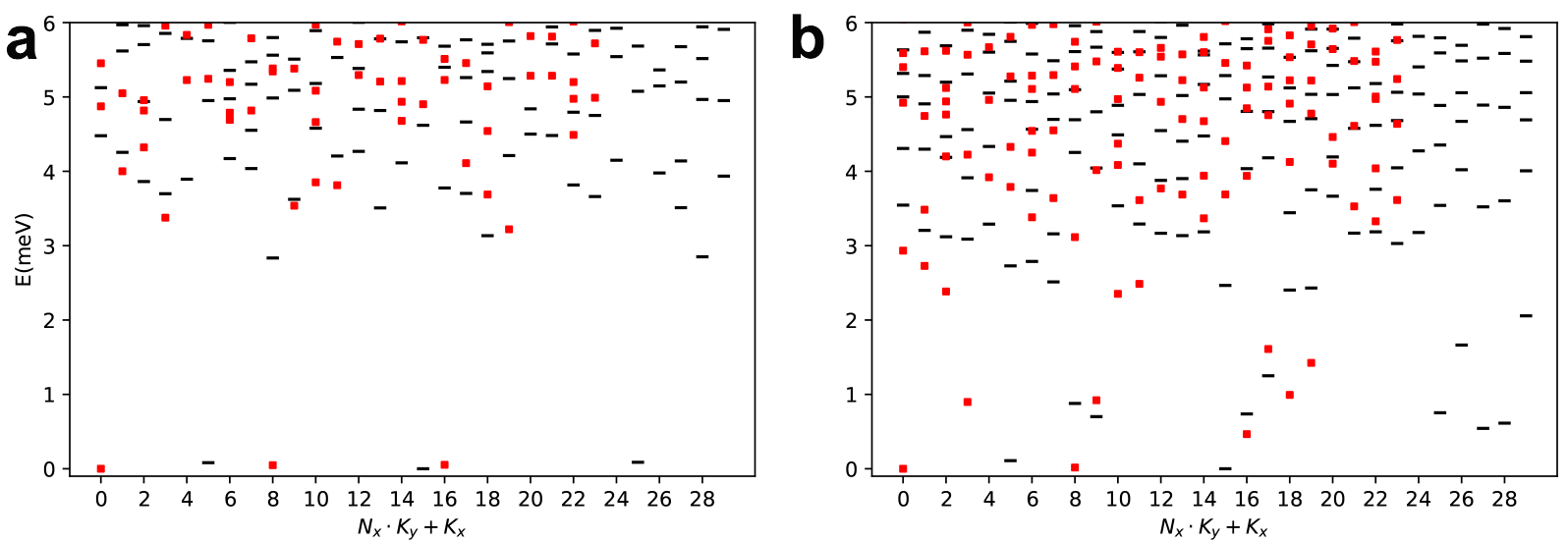}
		\caption{Result when consider larger system of $N_e=10,N_x\times N_y=5\times6$ (black) and mainly used size $N_e=8,N_x\times N_y=4\times6$ (red) of $\kappa=0.6$, (a) $\mathbf{k}_\phi=K_{\phi 1}$, (b)$ \mathbf{k}_\phi=G_{\phi }$.}
		\label{fig:largersystemcheck}
	\end{figure}
	In cases where the band gap is small, it is necessary to consider the influence of other bands. Generally, when multiple bands are taken into consideration, the Hilbert space will increase significantly, leading to high computational cost. Fortunately, by limiting the maximum or minimum number of filled electrons in remote bands, the diagonalization process can still be carried out \cite{Yu2024MultiED}. Our system requires a lattice as homogeneous as possible, which means that only even number of electrons $N_e=4,8,10,12 \cdots$ can be taken, otherwise, the ground state will not converge well. Consequently, with consideration on both efficiency and accuracy, we firstly take $N_x\times N_y=4 \times 3$ at $\nu=1/3$ filling. To check the convergence and validity of the system, we perform multi-band exact diagonalization according to the following workflow:

	1. Consider two flat band and force the upper flat band to fill at most one electron. The result is consistent with that when only the valence flat band is considered, indicating convergence, see Fig. \ref{fig:manybodygapconvergence} (a, b).
	
	2. Consider the valence flat band and the lower remote Dirac band and force the Dirac band to be in a fully filled state. The result is different from that when only the valence flat band is considered, see Fig. \ref{fig:manybodygapconvergence} (a, c).
	
	3. Consider the valence flat band and the lower remote Dirac band and force the Dirac band to be filled with one hole at most. The result is the same as that in case 2, indicating convergence , see Fig. \ref{fig:manybodygapconvergence} (c, d).
	
	These results indicate that the most ideal computation setting at  $4\times 3$ cases is the full filling of remote Dirac bands. However, this small cluster of $4\times3$ probably not convergence. Therefore, it should be further checked with larger lattices of $4\times 6$ for remote Dirac band effect. As an typical example from Fig. \ref{main-fig:edlocal}(a, f), as shown in Fig. \ref{fig:convegence}, the effect of Dirac band only slightly modify the many body spectrum. Therefore we can safely use single band projection results without any trouble.
	
	Additionally, we also compute larger system of $10$ electrons at $5\times6$ Lattice, the results also consist with $8$ electron cases. As depicted in Fig. \ref{fig:largersystemcheck}(a), the degenerated ground states are located at $K_x+K_y N_x=5,15,25$ for FCI states, while two fold degeneracy for CDW ground states located at $K_x+K_y N_x=5,15$.
	
	\section{Wannier Center evolution}
	\begin{figure}[b]
		\centering
		\includegraphics[width=1\linewidth]{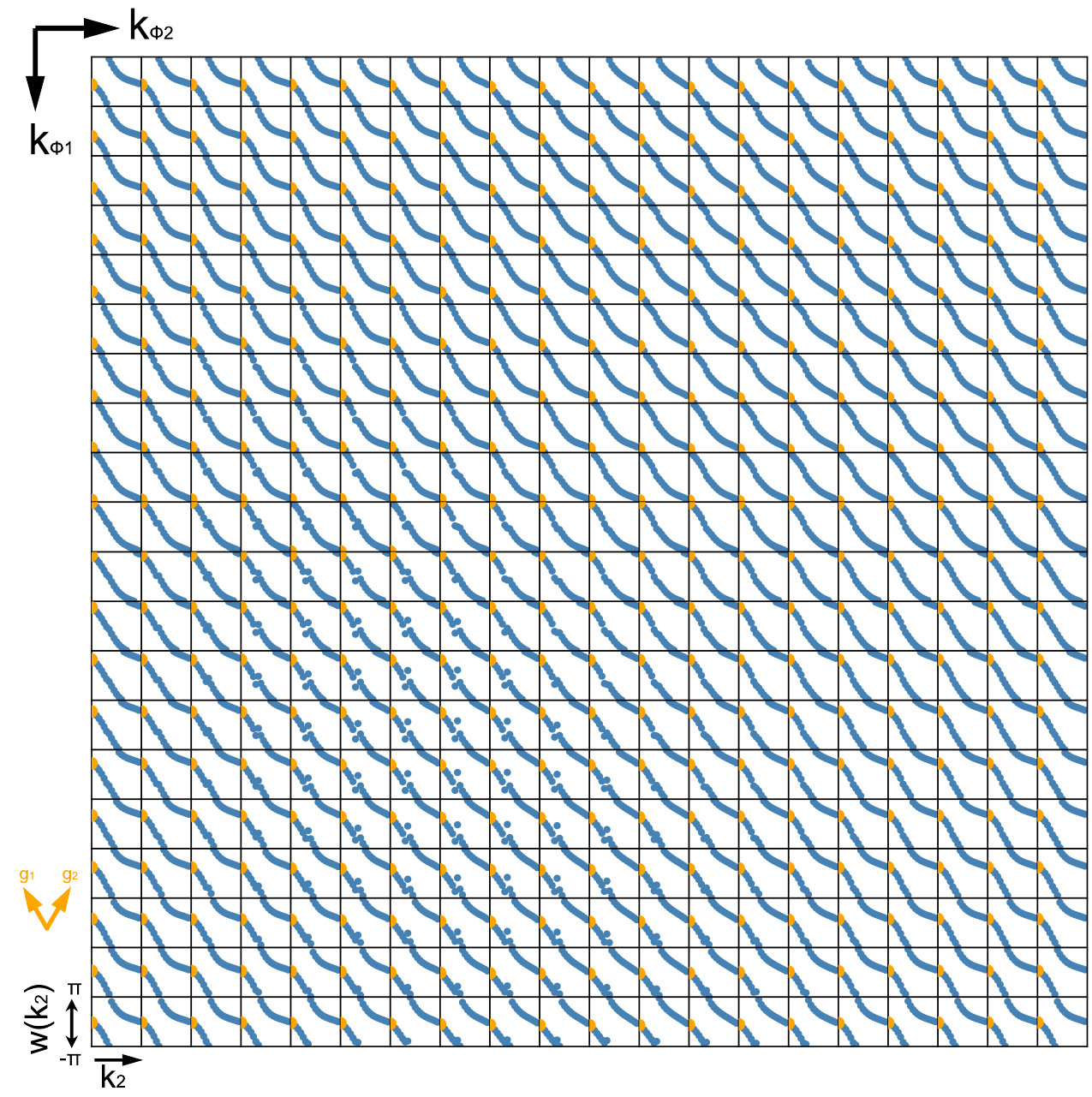}
		\caption{Wannier Center evolution in a supermoir\'e unit cell. Each square unit imply a local Wannier center shift along $k_2$. The orange dots are used as indicators to observe Wannier center evolution along $\mathbf{k}_\phi$ directions.}
		\label{fig:globalwanniershift}
	\end{figure}
	In the main text Fig. \ref{main-fig:ttgbandstructure} (g), we provide the Wannier center evolution diagram along $k_2$ and $k_{\phi1}$ direction by integrating out $k_1$ and a fixed $k_{\phi2}$. Here, we show the full evolution at supermoir\'e scale. As illustrate in Fig. \ref{fig:globalwanniershift}, the supermoir\'e unit cell is discrete into a $20\times20$ mesh, where each tiny square diagram denote the local Wannier center shift along $k_2$ in a one loop winding, which indicate $C=1$ of local flat bands. Meanwhile, Wannier center along $k_{\phi1}$ also shows one loop winding, indicating a topological Thouless pumping. As a comparison, there is no winding along $k_{\phi2}$ direction. The additional topological Thouless pumping making I-FCI introduced in the main text sharing different global properties with conventional FCI.

	\section{Dual lattice model derivation}
	
	In this section, we aims to formally relate the A-TTG continuum Hamiltonian to the four dimensional tight-biding (TB) Hamiltonian by momentum-coordinate duality.
	We begin with the Hamiltonian Eq. (\ref{main-eq:BM-mat}) near $-$K valley, rewriting it with single layer graphene TB Hamiltonian and into a more compact form,
	\begin{gather}
		H=t\sigma_+h (\mathbf{k})  +\tau_1T_1(\mathbf{r}) +\tau_2T_2(\mathbf{r}) + h.c.,
		\label{eq:TTG_Hamiltonian}
	\end{gather}
	where $t=2v_F/\sqrt{3} a$ is hopping energy of graphene, $2\sigma_+=\sigma_x+i\sigma_y$ represents sublattice degree of freedom, $h (\mathbf{k}) =\sum_{L=1}^3 \sum_{s=0}^2 e^{-i\mathbf{\Delta}_s \cdot(\mathbf{k+K}_L)}|L\rangle\langle L|$ and  $\Delta_0=-(\mathbf{a}_1+\mathbf{a}_2)/3,\Delta_1=\Delta_0+\mathbf{a}_1,\Delta_2=\Delta_0+\mathbf{a}_2$ with $\mathbf{a}_1=a[1,0],\mathbf{a}_2=a[1,\sqrt{3}]/2$ are primitive basis of graphene. We also denote the reciprocal basis of graphene as $\mathbf{b}_1,\mathbf{b}_j$, satisfied $\mathbf{a}_i\cdot\mathbf{b}_j=2\pi\delta_{ij}$ and $\mathbf{K}_L=\frac13(2 \mathbf{b}_1 +\mathbf{b}_2)+\Delta \mathbf{K}_L$ are -K points for each layer with $\Delta \mathbf{K}_1=q/3(\mathbf{g}_1+\mathbf{g}_2),\Delta \mathbf{K}_2=0,\Delta \mathbf{K}_3=p/3(\mathbf{g}_1+\mathbf{g}_2)$.   Note that, the graphene lattice basis are not exact. Here we use them as ultraviolet cut off to the continuum model, such that $a$ only need to be small enough. Therefore we have $\mathbf{b}_1= N\mathbf{g}_1-2N\mathbf{g}_2$ and  $\mathbf{b}_2= N\mathbf{g}_1+N\mathbf{g}_2$, where $N$ is the scaling parameter that renormalize $t$, $\mathbf{g}_j=k_\theta[\pm\sqrt{3},3]/2$ and $k_\theta=8\pi/3 a_0\sin(\theta_1/2)$ with $a_0=0.246$ $nm$.
	And the $\tau$ matrix imply the layer degree of freedom with $\tau_1=|1\rangle\langle 2|$ and $\tau_2=|3\rangle\langle 2|$.
	Remember $T_s=w(\kappa \sigma_0+\sigma_x\cos(2\pi s/3 )-\sigma_y\sin(2\pi s/3 ) )$ with $w=110$ $meV$ and $\kappa=0.5$, we can transform Eq. (\ref{eq:TTG_Hamiltonian}) into matrix form,
	\begin{align}
		H&=\sum_\mathbf{G}\sum_{I,s}\tau_{I}\otimes T_se^{i\delta \mathbf{q}_s \cdot \mathbf{R}}|\mathbf{G}\rangle\langle\mathbf{G+g}_{Is}| \label{eq:TTG_Hamiltonian2}
		\\\nonumber
		&+ t\sum_{s=0}^2  L_s \otimes \sigma_+ e^{-i\mathbf{\Delta}_s \cdot(\mathbf{k'+G})} |\mathbf{G}\rangle\langle \mathbf{G}|+h.c.  		
	\end{align}
	where $\mathbf{k=k'+G}$, $\mathbf{G}=n_1\mathbf{g}_1+n_2\mathbf{g}_2$, and
	$
		L_s=\sum_{L=1}^3 e^{-i\mathbf{\Delta}_s \cdot \mathbf{K}_L} |L\rangle\langle L|.\nonumber
	$
	We let $\delta \mathbf{q}_s \cdot \mathbf{R}= k_{\phi_s}$ for simplicity. Note that $\delta \mathbf{q}_{1}+\delta \mathbf{q}_{2}=-\delta \mathbf{q}_{0}$ so only $\mathbf{k}_\phi=({k}_{\phi_1},{k}_{\phi_2})$ are independent. Also we rewrite the Bloch momentum $\mathbf{k'}=k_1'\mathbf{g}_1 +k'_2\mathbf{g}_2$, and use $\mathbf{k}=(k_1',k_2')$ with abuse of notation for simplicity. 
	\begin{figure}
		\centering
		\includegraphics[width=\linewidth]{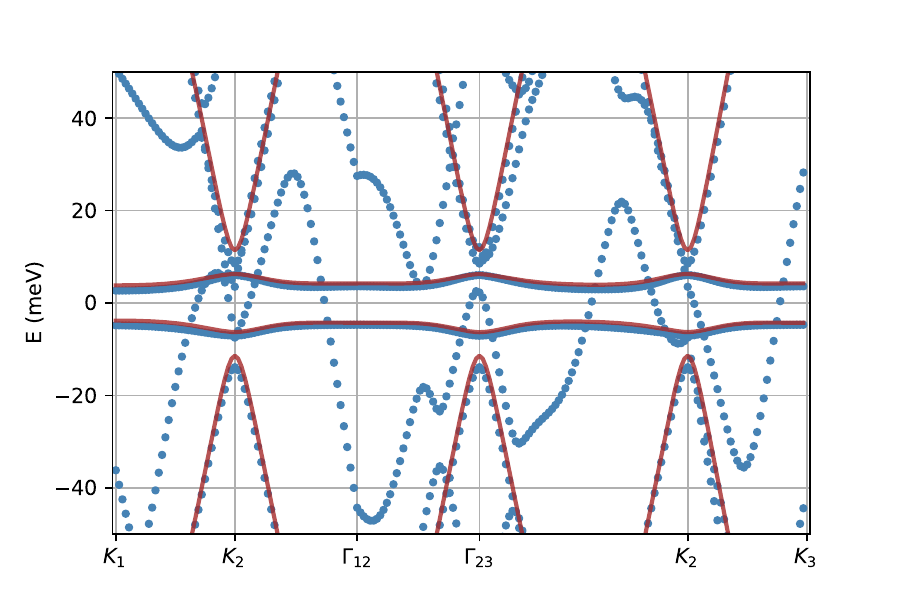}
		\caption{The comparison of resulting band structure between the original effective CM model (red) and dual model (dotted). }
		\label{fig:dualmodelband}
	\end{figure}

	This Hamiltonian can be written by introducing creation and annihilation operator $c^\dagger_{\mathbf{n,k},\gamma},c_{\mathbf{n,k},\gamma}$ where we write the sublattice and layer degree of freedom together as a new isospin degree of freedom labeled by $\gamma=1,\cdots,6$. Furthermore, it's more convenient to let $t_{I,s}=\tau_{I}\otimes T_s$ and $h_s=t L_s\otimes\sigma_+$. The scattering vectors are $\mathbf{g}_{Is}={e}_{Is,1}\mathbf{g}_1+{e}_{Is,2}\mathbf{g}_2$ with $\mathbf{e}_{Is} \in\mathbb{Z}\times\mathbb{Z}$. Therefore, the total interlayer coupling is expressed as,
	\begin{gather}
		H_T=\sum_{\mathbf{n},\gamma,\gamma'} \sum_{I,s} t^{\gamma\gamma'}_{I,s} e^{ik_{\phi s}}c^\dagger_{\mathbf{n+\mathbf{e}_{Is},k},\gamma}c_{\mathbf{n,k},\gamma'}+h.c. 
		\label{eq:InterLayerHami}
	\end{gather}
	For the intralayer part, denote $\mathbf{\Delta}_s\cdot \mathbf{g}_j=\alpha_{sj}$ with $j=1,2$. 
	\begin{gather}
		H_0=\sum_{\mathbf{n},\gamma,\gamma'}\sum_{s=0}^2 h_s^{\gamma\gamma'} e^{iF(\mathbf{n,k},s)}c^\dagger_{\mathbf{n,k},\gamma}c_{\mathbf{n,k},\gamma'}+h.c,
		\label{eq:IntraLayerHami}
	\end{gather}
	where, $F(\mathbf{n,k},s)=\alpha_{s1}(k_1+n_1)+\alpha_{s2}(k_2+n_2).$
	In this representation, if we only consider the interlayer term Eq. (\ref{eq:InterLayerHami}), the phase factor $\mathbf{k}_\phi$ only plays role as a pure gauge term, and have no influence to the eigenvalue of Eq. (\ref{eq:InterLayerHami}). However, it is not the case when  Eq. (\ref{eq:IntraLayerHami}) is involved. By a isospin-dependent gauge transformation to the operator, $c_{\mathbf{n,k},\gamma'}\rightarrow c_{\mathbf{n,k},\gamma'}e^{-i\theta_{\gamma'}}$, we can annihilate $k_{\phi s}$ in  Eq. (\ref{eq:InterLayerHami}) by define $k_{\phi s}=\theta_{\gamma'}-\theta_\gamma$. We then have an additional phase factor $-k_{\phi s}$ in Eq. (\ref{eq:IntraLayerHami}). After that, considering define $k_{ms}=\boldsymbol{\alpha}_s\cdot\mathbf{k}-k_{\phi s}\in[0,2\pi )$, we have the total Hamiltonian,
	\begin{align}
		H(\mathbf{k,k}_\phi)&=\sum_{\gamma,\gamma'=1}^6\sum_{s=0}^2\sum_{\mathbf{n}}\left(\sum_{I=1}^2 t^{\gamma\gamma'}_{I,s} c^\dagger_{\mathbf{n+\mathbf{e}_{Is},k},a}c_{\mathbf{n,k},b}\right)
		\nonumber\\
		 &+h_s^{\gamma\gamma'} e^{i\left(\boldsymbol{\alpha}_s\cdot\mathbf{n}+k_{ms}\right)}c^\dagger_{\mathbf{n,k},\gamma}c_{\mathbf{n,k},\gamma'}+h.c
		 \label{eq:totHami}
	\end{align}
	If we consider the global Hamiltonian with $G(\mathbf{k}_\phi)=1$
	\begin{align}
		H=\int &d\mathbf{k}_\phi \sum_\mathbf{k} H(\mathbf{k}_\phi,\mathbf{k})
		 = A \sum_{\mathbf{k}_m} H(\mathbf{k}_m)
		\nonumber\\
		 =A& \sum_{\gamma,\gamma'=1}^6\sum_{s=0}^2\sum_{\mathbf{n}}\sum_{\mathbf{k}_{m}}\left(\sum_{I=1}^2 t^{\gamma\gamma'}_{I,s} c^\dagger_{\mathbf{n+\mathbf{e}_{Is}},\mathbf{k}_{m},\gamma}c_{\mathbf{n},\mathbf{k}_{m},\gamma'}\right)
		 \nonumber\\
		 &+h_s^{\gamma\gamma'} e^{i\left(\boldsymbol{\alpha}_s\cdot\mathbf{n}+k_{m s}\right)}c^\dagger_{\mathbf{n},\mathbf{k}_{m},\gamma}c_{\mathbf{n},\mathbf{k}_{m},\gamma'}+h.c 
		 \label{eq:GtotHami}
	\end{align}
	where $A$ is a constant. If we view $\mathbf{k}_m$ as the momentum in the fiction 2D lattice $\{\mathbf{m}\}$, we can apply the Fourier transformation to the basis,
	\begin{gather}
		c_{\mathbf{n},k_{ms},\gamma}^\dagger=\frac{1}{N_m}\sum_{\mathbf{m}}e^{i\mathbf{k}_{m}\cdot \mathbf{m}} c_{\mathbf{n},\mathbf{m},\gamma}^\dagger 
	\end{gather}
	Neglect the constant $A$, we have a 4D dual lattice model on ($\mathbf{n},\mathbf{m}$)$\in \{\mathbf{n}\}\times\{\mathbf{m}\}$ ,
	\begin{align}
		H_\mathrm{4D}
		&=\sum_{\gamma,\gamma'=1}^6\sum_{s=0}^2\sum_{\mathbf{n,m}}\left(\sum_{I=1}^2 t^{\gamma\gamma'}_{I,s} c^\dagger_{\mathbf{n+\mathbf{e}_{Is}},\mathbf{m},\gamma}c_{\mathbf{n},\mathbf{m},\gamma'}\right)
		\nonumber\\
		&+h_s^{\gamma\gamma'} e^{i\boldsymbol{\alpha}_s\cdot\mathbf{n}}c^\dagger_{\mathbf{n},\mathbf{m+1}_s,\gamma}c_{\mathbf{n},\mathbf{m},\gamma'}+h.c 
		\label{eq:Dualmodel}
	\end{align}
	where $\mathbf{1}_{0,1,2}=[0,0]_\mathbf{m},[1,0]_\mathbf{m},[0,1]_\mathbf{m}$ and $\mathbf{e}_{I 0,1,2}=[0,0]_\mathbf{n},[I,0]_\mathbf{n},[0,I]_\mathbf{n}$. Eq. (\ref{eq:Dualmodel}) describe the quantum dynamic of the electron under magnetic flux on the 4D lattice, where the vector potential is written as $\mathbf{A}=-[0,0,\alpha_{11}x+\alpha_{12}y,\alpha_{21}x+\alpha_{22}y]$
	with $(\mathbf{n,m})$ represent lattice in 4D space labeled by $(x,y,z,w)$.
	\begin{figure}[b]
		\centering
		\includegraphics[width=\linewidth]{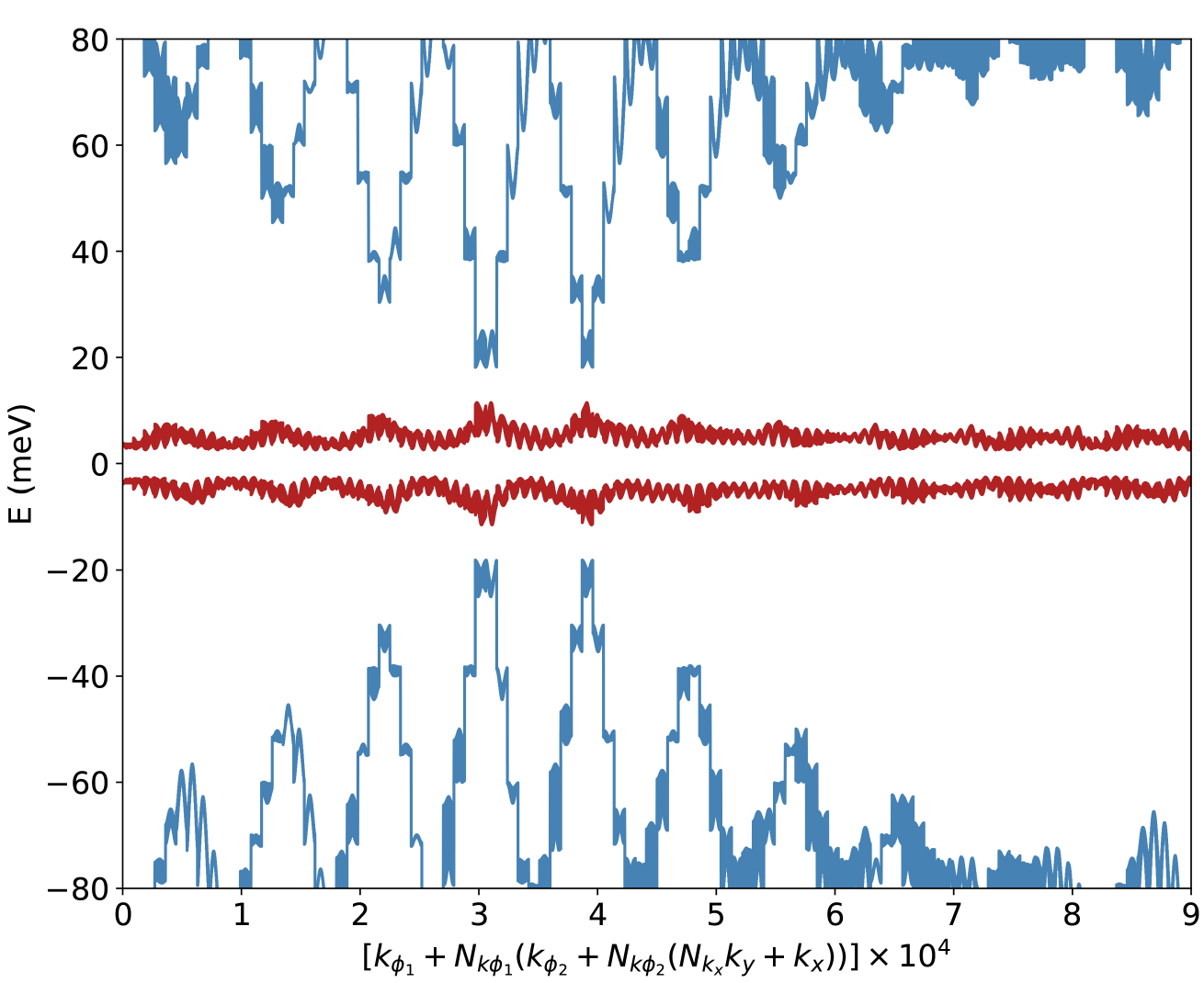}
		\caption{Band structure across 4D torus on $10\times10\times30\times30$ mesh.}
		\label{fig:band4d}
	\end{figure}
	As illustrated in Fig. \ref{fig:dualmodelband}, the continuum model described by Eq. (\ref{main-eq:BM-mat}) can be interpreted as an effective model of the dual model. Due to the intrinsic anomaly associated with the $\sigma \cdot \mathbf{k}$ term, Eq. (\ref{main-eq:BM-mat}) cannot be entirely equivalent to a lattice model. To compensate for this anomaly, additional nonphysical states must be introduced by modifying $\sigma \cdot \mathbf{k}$ to $h(\mathbf{k})$. However, these states can be projected out, as the emergent valley $U(1)$ symmetry remains intact when $N$ is large.
	
	\section{Calculation of the Second Chern number}
	The numerical calculation of the second Chern number $C_2$ was performed on the $N_x\times N_y\times N_{k_{\phi1}}\times N_{k_{\phi2}}=10\times10\times30\times30$ mesh in a 4D torus for fast convergence, which is defined as,
	\begin{equation}
		C_2 = \frac{1}{(2\pi)^2} \int_{\mathbb{T}^4} d^4k \, \epsilon^{\mu\nu\rho\sigma} \langle \partial_\mu u_k | \partial_\nu u_k \rangle \langle \partial_\rho u_k | \partial_\sigma u_k \rangle,
	\end{equation}
	where $\epsilon^{\mu\nu\rho\sigma}$ is the levi-civita tensor, $k=(\mathbf{k},\mathbf{k}_\phi)$ parameterize the dual 4D momentum and $\partial_{\mu}=\frac{\partial}{\partial k^\mu}$. Efficient algorithm was used to perform this calculation, see Refs. \cite{Mochol2018SecondChern,YJXiu2023SecondChern} for details.  The band structure on this 4D torus is illustrated in Fig. \ref{fig:band4d}.

	\section{Effect of dielectric constant }
	\begin{figure*}
		\centering
		\includegraphics[width=1\linewidth]{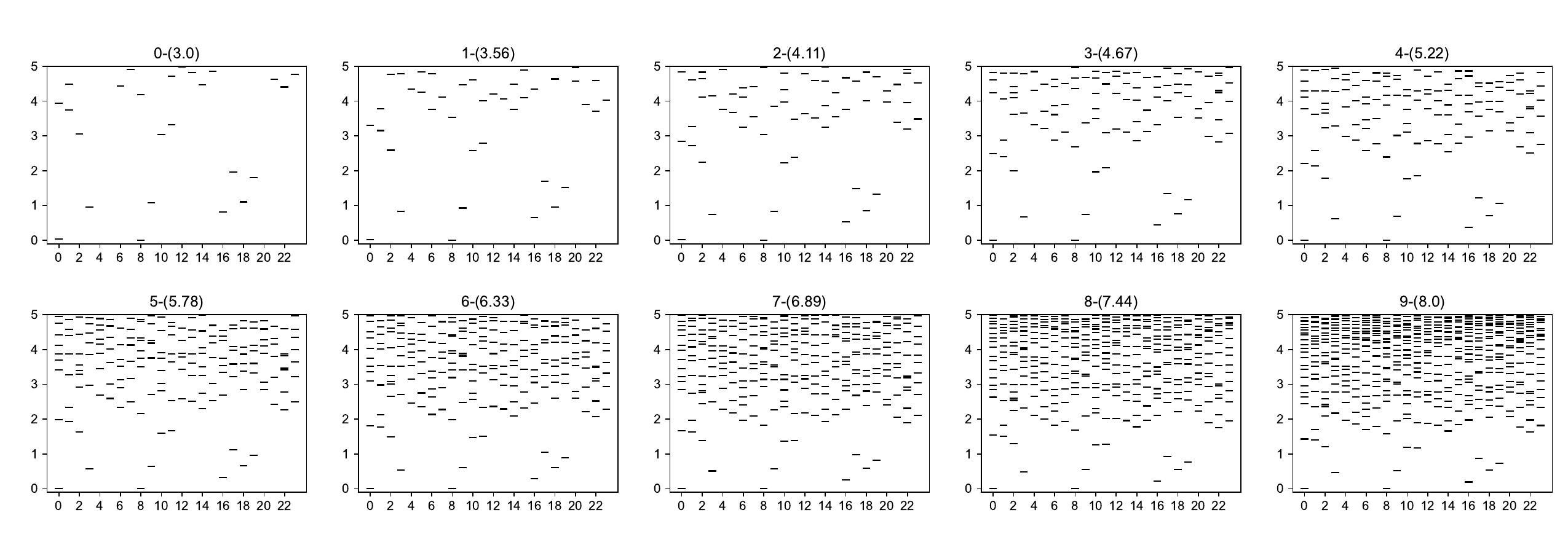}
		\includegraphics[width=1\linewidth]{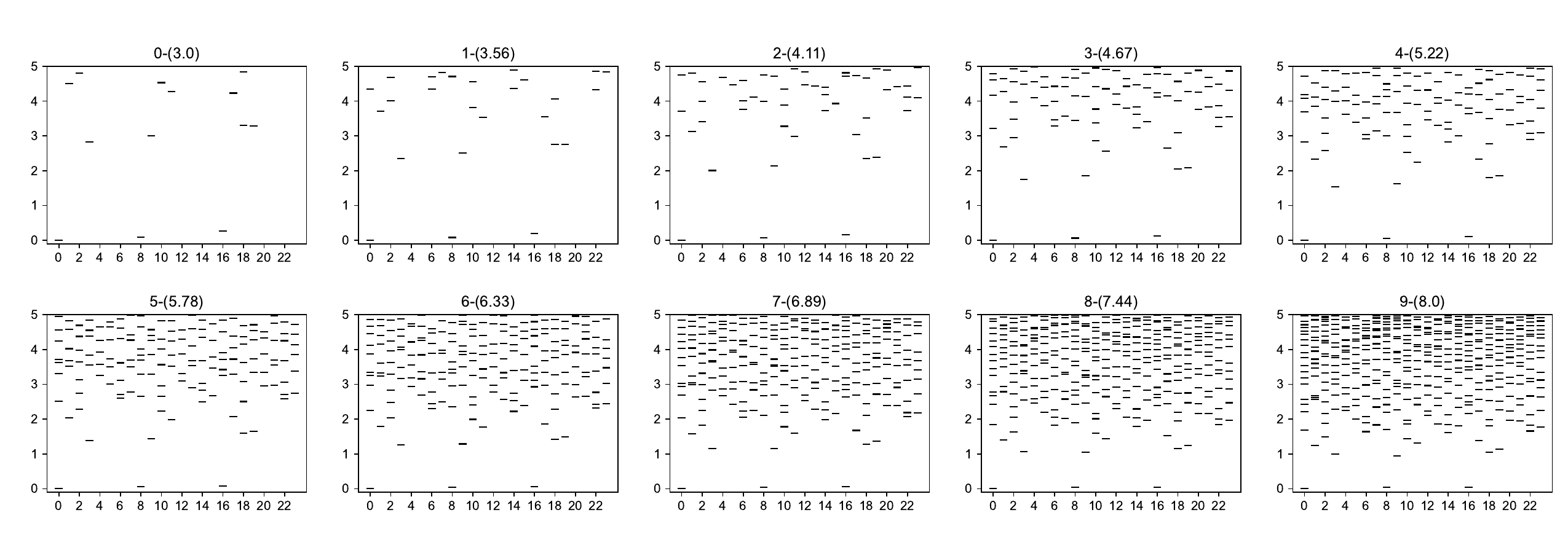}
		\caption{The first two rows gives ED spectrum at $\Gamma_\phi$ while the last two rows is at $K_{\phi 1}$}
		\label{fig:-edresultsepsg}
	\end{figure*}
	In practice, the intensity of Coulomb interaction will greatly influence by the dielectic constant where we set to be $\varepsilon=4\varepsilon_0$ in the main text. Here we vary $\varepsilon_r=\varepsilon/\varepsilon_0$ from $3$ to $8$. As illustrate in Fig. \ref{fig:-edresultsepsg}, the increasing $\epsilon_r$ decrease the many body gap of the system. At $\mathbf{k}_\phi=K_{\phi1}$, the FCI ground states are always maintain. While at $\mathbf{k}_\phi=\Gamma_{\phi}$, the state at $\mathbf{K}=(4,0)$ is lowering. As we consider $\epsilon_r\le6$ the states are in the CDW phases while $\epsilon_r>6$, it turns to FCI states. This out come is interesting, for FCI traditionally need stronger interaction, however, in this cases, things become abnormal instead. As a conclusion, the increasing $\epsilon$ do not modified the system heavily, indicating the relative robustness of FCI states.
	\section{Truly commensurate case at $\theta_1=\theta_2$}
\begin{figure*}
		\centering
		\includegraphics[width=1\linewidth]{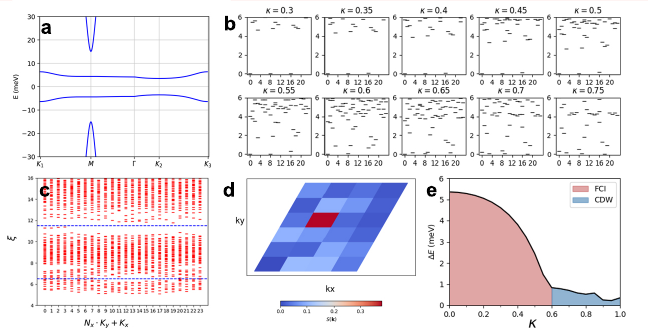}
		\caption{
			(a) band structure at $\theta_1==\theta_2=1.71^\circ$ when $\kappa=0.6$.
			(b) ED results are given below with respect to $K_x+N_xK_y$ at different $\kappa$.
			(c) The PES with respect to $K_x+N_xK_y$ at phase transition points ( $\kappa=0.6$ ) depicts mixture properties of FCI (The upper blue line with 1088 states below) and CDW (The lower blue line with  168 states below).
			(d) CDW nature can also be checked by structural factor $S(\mathbf{k})=\langle\rho_\mathbf{k}\rho_{-\mathbf{k}}\rangle-\frac{1}{N_k}\delta_{\mathbf{k},0}$  at $\kappa=0.6$ for an obvious peak exists.
			(e) ED results on FCI gap for $\theta_1=\theta_2\approx1.71^\circ$at different $\kappa$.
			}
		\label{fig:eqangledata}
	\end{figure*}

	As illustrate in Fig. \ref{fig:eqangledata}, ED and PES calculations are performed at $\nu=1/3$ for truly commensurate physical  case (using the magic angles $\theta_1=\theta_2\approx1.71^\circ$) under the same parameters (Dirac velocity, interlayer couplings, substrate potentials, and corrugation $\kappa$) as in the main text. Although an FCI phase does emerge in this commensurate configuration, it is less stable than that observed in the main text. For example, at $\kappa=0.6$, the configuration with $\theta_2=2\theta_1=2.8^\circ$ exhibits local FCI phases away from $\mathbf{k}_\phi=0$ and charge density wave (CDW) phases at $\mathbf{k}_\phi=0$ (See Fig. \ref{main-fig:edkappa}(a) yellow line), whereas the configuration with $\theta_1=\theta_2\approx1.71^\circ$ shows a global  CDW phase with an ED spectrum very similar to the local CDW regions in the main text. When $\kappa$ is small, the behavior is analogous, with the FCI gap emerging similarly to the $\mathbf{k}_\phi=0$ as a nearby commensurate system in our work. As a result, this consequence also validate incommensuration is curial to stabilize FCI states.  

\clearpage
\bibliography{TTG_FCI_refs.bib}